\documentclass[fleqn,usenatbib]{mnras}

\usepackage{newtxtext,newtxmath}

\usepackage[T1]{fontenc}
\usepackage{ae,aecompl}

\usepackage{graphicx}	
\usepackage{amsmath}	
\usepackage{amssymb}	
\usepackage{hyperref}
\usepackage[]{units}

\newcommand{\msun}{M_{\sun}}
\newcommand{\oft}{\left(t\right)}
\defcitealias{grudic:2016.sfe}{G18}
\defcitealias{lee:2016.gmc.eff}{L+16}
\defcitealias{vuti:2016.gmcs}{V+16}
\defcitealias{heyer:2016.clumps}{H+16}
\defcitealias{wu:2010.clumps}{W+10}
\defcitealias{evans:2014.sfe}{E+14}
\defcitealias{lada:2010.gmcs}{L+10}

\title[Scatter in the SFE of Molecular Clouds]{On The Nature of Variations in the Measured Star Formation Efficiency of Molecular Clouds}

\author[Grudi\'{c} et al.]{Michael Y. Grudi\'{c}$^{1}$\thanks{mgrudich@caltech.edu}, Philip F. Hopkins$^{1}$, Eve J. Lee$^{1,2}$, Norman Murray$^{3,4}$, \newauthor Claude-Andr\'{e} Faucher-Gigu\`{e}re$^{5}$,  L. Clifton Johnson$^{5}$
 \\
$^{1}$TAPIR, Mailcode 350-17, California Institute of Technology, Pasadena, CA 91125, USA\\
$^{2}$Walter Burke Institute for Theoretical Physics, Pasadena, CA 91125, USA\\
$^{3}$Canadian Institute for Theoretical Astrophysics, 60 St. George Street, University of Toronto, ON M5S 3H8, Canada\\
$^{4}$Canada Research Chair in Astrophysics \\
$^{5}$Department of Physics and Astronomy and CIERA, Northwestern University, 2145 Sheridan Road, Evanston, IL 60208, USA
}

\pubyear{2018}

\begin{document}
\label{firstpage}
\pagerange{\pageref{firstpage}--\pageref{lastpage}}
\maketitle

\begin{abstract}
Measurements of the star formation efficiency (SFE) of giant molecular clouds (GMCs) in the Milky Way generally show a large scatter, which could be intrinsic or observational. We use magnetohydrodynamic simulations of GMCs (including feedback) to forward-model the relationship between the true GMC SFE and observational proxies. We show that individual GMCs trace broad ranges of observed SFE throughout collapse, star formation, and disruption. Low measured SFEs ($\ll 1\%$) are ``real'' but correspond to early stages; the true ``per-freefall'' SFE where most stars actually form can be much larger. Very high ($\gg 10\%$) values are often artificially enhanced by rapid gas dispersal. Simulations including stellar feedback reproduce observed GMC-scale SFEs, but simulations without feedback produce $20\times$ larger SFEs. Radiative feedback dominates among mechanisms simulated. An anticorrelation of SFE with cloud mass is shown to be an observational artifact. We also explore individual dense ``clumps'' within GMCs and show that (with feedback) their bulk properties agree well with observations. Predicted SFEs within the dense clumps are $\sim2\times$ larger than observed, possibly indicating physics other than feedback from massive (main sequence) stars is needed to regulate their collapse.
\end{abstract}



\section{Introduction}
\label{section:intro}





Giant molecular clouds (GMCs) are the sites of star formation within the Galaxy \citep{myers:1986.gmcs,shu:1987.review, scoville:1989.gmcs}. They are regions of elevated ($>\unit[100]{cm^{-3}}$) molecular gas density with typical masses $M \sim \unit[4\times10^4-4\times10^6]{\msun}$ and radii $R \sim \unit[10-100]{pc}$, with a characteristic surface density on the order of $\Sigma \sim  \unit[100]{\msun\,pc^{-2}}$ in local galaxies \citep{solomon:gmc.scalings,bolatto:2008.gmc.properties}. Star-forming GMCs tend to host massive stars and HII regions, have supersonically-turbulent internal gas motions \citep{larson:gmc.scalings}, and may be self-gravitating \citep{mckee.tan:2003}. It is thus believed that the evolution of star-forming clouds is the result of a complex interplay of stellar and protostellar feedback, supersonic magnetohydrodynamic (MHD) turbulence, and gravity in concert \citep{mckee:2007.review,krumholz:2014.feedback.review}.

Possibly the most powerful diagnostic of the effects of these physical mechanisms is the {\it star formation efficiency} (SFE) of a molecular cloud: the fraction of the molecular gas mass converted to stars. The question of what fraction of a molecular cloud's mass is converted to stars, and how quickly, is a fundamental one in star formation theory. Turbulence, magnetic fields, and feedback can all oppose the gravitational collapse that leads to star formation, and in doing so they can reduce the SFE to varying degrees. 

The SFE of star-forming clouds has been measured with many different methodologies and tracers of both stellar mass and molecular gas mass\footnote{See \S\ref{section:sfe.review} for the various definitions of SFE and how they are measured; our discussion thus far is not specific to any in particular.}, but virtually all studies of Local Group clouds have found that typical (i.e. median) values are on the order of $1\%$ \citep{myers:1986.gmcs,mooney.solomon:1988,williams.mckee:1997, evans:2009.sfe,lada:2010.gmcs,heiderman:2010.gmcs,murray:2010.sfe.mw.gmc,lee:2016.gmc.eff,vuti:2016.gmcs,ochsendorf:2017.gmcs}. However, the cited studies have also generally found considerable {\it scatter} in the SFE -- typically at least $0.5\,{\rm dex}$. 

If this scatter reflects an actual diversity in the intrinsic scale of the SFE of molecular clouds with otherwise similar properties, then it presents a serious challenge to theories that attempt to explain the SFE of molecular clouds in terms of their large-scale turbulent properties such as the virial parameter or Mach number \citep[e.g.][]{km2005, pn2011, hc2011, federrath:2012.sfr.vs.model.turb.boxes}, as the variations in these properties were found to account for less than $\sim 0.24\,{\rm dex}$ of the observed scatter (\citet{lee:2016.gmc.eff}). It would also challenge theories that attempt to explain the SFE of molecular clouds in terms of the balance of feedback from massive stars and gravity in a collapse-blowout scenario \citep{fall:2010.sf.eff.vs.surfacedensity,murray:molcloud.disrupt.by.rad.pressure,dale:2012, dale:2014,hopkins:fb.ism.prop,2014MNRAS.439.3420M,raskutti:2016.gmcs,kim:2017.rhd,grudic:2016.sfe}: stochastic variations in the SFE for a fixed set of cloud parameters tend to be rather small in numerical simulations, and the SFE spread predicted from the variations in the cloud properties that determine the SFE (e.g. surface density for momentum-conserving feedback, escape velocity for expanding HII bubbles) are also much smaller than observed.
However, a more likely explanation for the scatter comes from the fact that these observationally-inferred efficiencies have intrinsic variation over the lifetime of a GMC that does not necessarily reflect the true SFE. Although one  theoretical picture of star-forming clouds is one of quasi-equilibrium \citep{zuckerman1974,shu:1987.review,krumholz.matzner.mckee:2006} with a relatively steady star formation rate (SFR), the age distributions of nearby star-forming regions suggest an accelerating SFR \citep{palla2000}. This can produce large variations in the stellar mass tracer over the lifetime of a single cloud. Such acceleration has a theoretical basis in the behaviour of self-gravitating isothermal supersonic turbulence, wherein it is expected that $\dot{M}_\star \propto t$ from both analytic considerations \citep{murray.chang:2015} and hydrodynamics simulations \citep{lee:2015.gravoturbulence, vazquez:2015,murray:2017}. 

The inferred gas mass will also vary over the lifetime of the cloud: while the effect of gas consumption might be negligible if overall efficiencies are small, molecular gas will also be destroyed and ejected by stellar feedback, so toward the end of a cloud's star-forming lifetime the inferred efficiency might be biased upward. Semi-analytic models of cloud evolution that model both the effects of time-varying SFR and mass loss due to feedback have been found to produce SFE scatter similar to that observed \citep{feldmann:2011,lee:2016.gmc.eff}. Meanwhile, many simulations of star-forming clouds have been done that consider at least some subset of the important stellar feedback channels (\citealt{murray:molcloud.disrupt.by.rad.pressure, vazquez:2010, dale:2012, dale:2013.momwinds, colin:2013, dale:2014, skinner:2015.ir.molcloud.disrupt, raskutti:2016.gmcs,howard:2016, howard:2017,vazquez:2017, dale:2017, kim:2017.rhd, gavagnin:2017.rhd.cluster.formation,grudic:2016.sfe,kim:2018}, for review see \citealt{krumholz:2014.feedback.review,dale:2015.review}), however of these only \citet{geen:2017} has addressed the specific problem of the interpretation of predicted molecular cloud SFEs vis-a-vis observations. They performed MHD simulations of the evolution of a low-mass molecular cloud with ionizing radiation feedback, and compared simulation results with SFE measurements in nearby star-forming regions derived from YSO counts \citep[e.g.][]{lada:2010.gmcs}. Notably, they found that errors and biases in the inferred SFE can be quite large depending upon the time of observation during the cloud lifetime. The also found that the stellar feedback was necessary to reproduce observations, finding that observations were most consistent with simulated clouds of mean surface density $\Sigma_{gas} \sim 40\,M_\odot{\rm pc^{-2}}$. 

In this paper we use a suite of MHD cloud collapse simulations modeling Milky Way GMCs to make a self-consistent prediction for the evolution of the SFE-related observables of a star-forming cloud. By including the effects of feedback from massive stars (in the form of stellar winds, radiation, and supernova explosions), we are able to follow the entire cloud lifetime from initial collapse to eventual disruption by stellar feedback. We will show the SFE observations are reasonably consistent with the hypothesis that GMCs with a given set of bulk properties do not have widely different star formation histories -- rather, the observed spread in SFE is comparable to that observed over the lifetime of a single cloud. The model of feedback-moderated star formation can thus explain the observed SFEs of molecular clouds.

\section{Star Formation Efficiency in Theory and Observation}

\label{section:sfe.review}
\begin{table}
\begin{tabular}{l|l|l}
Symbol & Name & Definition \\ \hline
$\epsilon_{int}$ & Integrated SFE & Eq. \ref{eq:epsint} \\
$\epsilon$ & Instantaneous SFE &  Eq. \ref{eq:eps} \\
$\epsilon_{ff}$ & Per-freefall SFE & Eq. \ref{eq:eff} \\
$\epsilon_{obs}$ & Tracer-inferred instantaneous SFE & Eq. \ref{eq:epsobs} \\
$\epsilon_{ff,obs}$ & Tracer-inferred per-freefall SFE & Eq. \ref{eq:effobs} \\
\end{tabular}
\caption{Summary of the various concepts of star formation efficiency discussed in this paper, with defining equations in Section \ref{section:sfe.review}.}
\label{table:sfe.review}
\end{table}

\begin{table*}
\begin{tabular}{l|l|l|l|l|l|l|l|l}
Study & Class & $M_{gas}$ Tracer & $M_\star$ Tracer & $\log \frac{n_{{\rm H}_2}}{{\rm cm}^{-3}}$ & $\log \frac{\Sigma_{gas}}{\msun\,{\rm pc}^{-2}}$ & $\log \epsilon_{obs}$ & $\log \epsilon_{ff,obs}$ \\ \hline
\citet{wu:2010.clumps} & Dense clumps & HCN $1\rightarrow0$ & FIR & $4.08_{3.38}^{4.68}$ & $3.00_{2.63}^{3.38}$ &  $-1.10_{-1.76}^{-0.86}$ & $-1.44_{-2.03}^{-0.82}$  \vspace{0.1cm} \\
\citet{evans:2014.sfe} & GMCs & Dust Extinction & YSO counts & $2.62_{2.33}^{2.99}$ & $1.86_{1.76}^{1.97}$ & $-1.72_{-2.19}^{-1.37}$ & $-1.79_{-2.38}^{-1.48}$    \vspace{0.1cm} \\
\citet{heyer:2016.clumps} & Dense clumps & Dust Emission  & YSO counts & $4.11_{3.64}^{4.62}$ & $2.78_{2.56}^{3.04}$ & $-1.11_{-1.66}^{-0.62}$ &  $-1.32_{-1.87}^{-0.84}$ \vspace{1mm} \\
\citet{vuti:2016.gmcs} & GMCs & $^{13}{\rm CO}$ $1\rightarrow0$ & Free-free, MIR & $2.20_{1.69}^{2.72}$ & $1.94_{1.67}^{2.23}$ & $-2.26^{-1.82}_{-2.94}$ & $-2.40_{-2.89}^{-1.99}$  \vspace{1mm} \\
\citet{lee:2016.gmc.eff} & GMCs & $^{12}{\rm CO}$ $1\rightarrow0$ & Free-free & $1.36_{0.78}^{1.82}$ &  $1.88_{1.40}^{2.19}$ & $-1.97_{-2.76}^{-1.20}$ & $-1.73_{-2.66}^{-0.96}$ \\
\end{tabular}
\label{table:observations}
\caption{Methodologies, parameter space, and summarized SFE results of several recent studies of star-forming clouds or clumps in the Milky Way. All quantities are given in the format ${\rm median}^{+\sigma}_{-\sigma}$. {\it Class}: class of star-forming cloud studied: GMC or dense clump. {\it $M_{mol}$ Tracer}: method used to obtain the properties of the molecular gas distribution. {\it $M_{\star}$ Tracer}: Emission type or object count used to estimate the SFR or stellar mass present. {\it $n_{{\rm H}_2}$}: number density of molecular hydrogen at the volume-averaged cloud density $3M/\left(4\pi R^3\right)$, in ${\rm cm}^{-3}$. $\Sigma_{gas}$: of the mean gas surface density $M/\pi R^2$. $\epsilon_{obs}$: observed instantaneous SFE (Eq. \ref{eq:epsobs}). $\epsilon_{ff,obs}$: Observed per-freefall SFE (Eq. \ref{eq:effobs}).}
\end{table*}

There are several flavours of star formation efficiency, some of which are motivated by observational convenience and others which are motivated by theory. First, we emphasize that we are interested in SFE as on the scale of individual gas clouds, rather than the SFE integrated over a larger region or an entire galaxy. In principle, these two SFEs can be completely decoupled from one another, and in feedback-regulated models for the \citet{kennicutt98} relation, they generally are to some extent \citep{thompson:rad.pressure,ostriker.shetty:2011,cafg:sf.fb.reg.kslaw,orr:2018.kennicutt.schmidt}.

Among the possible cloud-scale SFEs, the most conceptually straightforward is the ``integrated'' star formation efficiency, the fraction of the gas mass that is converted to stars across the entire lifetime of a cloud:
\begin{equation}
\epsilon_{int}=\frac{M_\star\left(t=\infty\right)}{M_{gas}\left(t=0\right)}, \label{eq:epsint}
\end{equation}
where $M_\star$ is the total mass in stars formed and $M_{gas}$ is the total gas mass. $\epsilon_{int}$ is of particular interest because it is sensitive to the details of stellar feedback physics, as eventually a sufficient number of massive stars will form to expel the gas. $\epsilon_{int}$ ultimately determines the mapping between the GMC and star cluster mass functions in a galaxy, 
and dictates the fraction of stars remaining in a gravitationally-bound clusters after gas expulsion
\citep{tutukov:1978,hills:1980,mathieu:1983,lada:1984,1985ApJ...294..523E,baumgardt.kroupa:2007}.
Although this quantity is ubiquitously reported in numerical simulations of star-forming clouds, it is not readily observable. It is difficult to define an unambiguous ``initial'' gas mass because
the evolution of the mass of
GMCs 
is often highly dynamic, subject to ongoing processes of accretion, merging and splitting \citep{dobbs:2013}. However, even supposing that a completely isolated, self-gravitating initial gas mass can be identified, the desired gas and stellar masses must be measured at the beginning and the end of the star-forming lifetime respectively. Thus, although $\epsilon_{int}$ is a quantity of great theoretical interest, we must resort to measuring it by proxy.

Instead, one might measure the ``instantaneous" star formation efficiency, which is simply the mass fraction of stars associated with the star-forming cloud at a given time \citep{myers:1986.gmcs}:
\begin{equation}
\epsilon = \frac{M_\star\left(t\right)}{M_\star \left(t\right) + M_{gas} \left(t\right)}. \label{eq:eps}
\end{equation}
$\epsilon$ will evolve from 0 to some finite value during the star-forming lifetime of a cloud, so a certain amount of scatter in this quantity is expected even for a population of clouds with identical properties. As $t\rightarrow \infty$, $\epsilon \rightarrow \epsilon_{int}$.

The natural timescale for the evolution of self-gravitating objects is the gravitational free-fall time, 
\begin{equation}
t_{ff} = \sqrt{\frac{3\pi}{32 G \bar{\rho}}} = \unit[13.2]{Myr}\left(\frac{\Sigma}{\unit[50]{\msun\,pc^{-2}}}\right)^{-\frac{1}{2}}\left(\frac{R}{\unit[100]{pc}}\right)^{\frac{1}{2}}, \label{eq:tff}
\end{equation}
where $\bar{\rho}$ is the volume-averaged density of the cloud, $\Sigma = M/\pi R^2$ is the average surface density, and $R$ the effective radius. This fact has motivated the development of theoretical models that predict the {\it per-freefall} SFE $\epsilon_{ff}$, the fraction of gas converted to stars per freefall time \citep{km2005, pn2011, hc2011, federrath:2012.sfr.vs.model.turb.boxes}:
\begin{equation}
\epsilon_{ff} = \frac{\dot{M}_\star \left(t\right) t_{ff}\left(t\right)}{M_{gas}\left(t\right)}, \label{eq:eff}
\end{equation}
where $\dot{M}_\star$ is the star formation rate (SFR).
These theories typically predict $\epsilon_{ff} \sim 1\%$ for molecular clouds with properties similar to those observed in local spiral galaxies \citep[e.g.][]{bolatto:2008.gmc.properties}, solely from the properties of isothermal supersonic turbulence plus a gravitational collapse criterion. Because these physics are scale-free, this could potentially explain the observation that $\epsilon_{ff} \sim 1\%$ on a wide range of scales from galaxies to dense star-forming clumps \citep{krumholz:2012.universal.sf.efficiency}. 

However, such a steady and universal SFE has not been found in hydrodynamics simulations of self-gravitating isothermal turbulence, with or without a source of turbulent stirring to maintain a constant virial parameter \citep{kritsuk:2011.density.pdf.power.law, padoan:2012.sfe, lee:2015.gravoturbulence,murray:2017}. Rather, simulations with virial parameters $\sim 1$ have found that $\epsilon_{ff}$ tends to increase roughly linearly to a saturation point on the order of several tens of percent. This saturation point has a residual dependence upon the magnetic field strength at the factor of 2 level \citep{federrath:2012.sfr.vs.model.turb.boxes}. The reason for this discrepancy is that in the presence of self-gravity, the density PDF deviates from the log-normal form assumed by the analytic theories, forming a high-density power-law tail. Such power-law tails have been observed in the extinction maps of star-forming clouds \citep{kainulainen:2009.density.pdf,lombardi:2014.density.pdf,schneider:2015a.power.law.tail,schneider:2015b.power.law.tail}. When this power-law tail is incorporated into analytic theory, the effect upon the SFE is captured more accurately \citep{burkhart:2018.density.pdf}.

\citet{grudic:2016.sfe} (hereafter \citetalias{grudic:2016.sfe}) argued that $\epsilon_{ff} \sim 1\%$ is the typical value observed for molecular clouds because feedback from massive stars is able to prevent runaway star formation, and that the ubiquity of the observed $1\%$ value is a consequence of the lack of variation of cloud surface density $\Sigma$, which they found determined both $\epsilon_{int}$ and $\epsilon_{ff}$. However, this cannot explain slow star formation in regions where massive stars are absent. \citet{federrath:2015.mhd.outflows} found that protostellar outflow feedback can bring $\epsilon_{ff}$ down to values on the order of $1\%$ in the regime of low-mass cluster formation, but this mechanism is unlikely to scale up to more massive systems. It is thus possible that the protostellar and massive stellar feedback complement each other in limiting the per-freefall SFE of molecular clouds on different scales.

Measuring $\epsilon$ and $\epsilon_{ff}$ requires some estimate of the stellar mass formed and the currently-present gas mass. We distinguish between the true instantaneous SFE $\epsilon$ and its observational tracer-inferred value:
\begin{align}
\epsilon_{obs} &= \frac{M_{\star,tr}\left(t\right)}{M_{\star,tr}\left(t\right) + M_{mol,tr}\left(t\right)}, \label{eq:epsobs}
\end{align}

where $M_{mol,tr}$ and $M_{\star, tr}$ are the tracer-inferred molecular gas and stellar masses, respectively. Similarly, we define the tracer-inferred proxy for $\epsilon_{ff}$:
\begin{equation}
\epsilon_{ff,obs} = \frac{M_{\star,tr}\oft\, t_{ff}\oft }{M_{mol,tr}\oft \,\tau_{tr}}, \label{eq:effobs}
\end{equation}
where we introduce the characteristic lifetime $\tau_{tr}$ of the species being traced, so that the SFR $\dot{M}_\star$, which is not directly observable, may be estimated as $M_{\star,tr}/\tau_{tr}$.

\subsection{Stellar Mass Tracers}

There are several methods for estimating $M_{\star}$. The most readily-measured tracer of stellar mass is the emission associated with HII regions, such as far IR \citep{myers:1986.gmcs}, mid IR \citep{vuti:2016.gmcs}, or free-free emission \citep{murray:2010.wmap,lee:2012.wmap,vuti:2016.gmcs}. This flux is dominated by the contribution from the reprocessed radiation from young, massive stars, and effectively traces the mass in stars younger than the ionizing flux-weighted mean stellar lifetime of a stellar population, $\tau_{MS}=\unit[3.9]{Myr}$ \citep{mckee:1997.ob,murray:2010.sfe.mw.gmc}. We refer to this stellar mass as the ``live'' stellar mass, $M_{\star,live}$. $M_{\star,live}$ can underestimate the total stellar mass formed in a cloud if its star formation history spans longer than $\tau_{MS}$, which appears to be the case for a majority of local GMCs \citep{2009ApJS..184....1K,2010ARA&A..48..547F,murray:2010.sfe.mw.gmc,lee:2016.gmc.eff}. 

Another tracer of the formed stellar mass is the mass of young stellar objects (YSOs), 
$M_{young}$,
which can be measured in sufficiently well-resolved star-forming clouds or clumps \citep{evans:2009.sfe,heiderman:2010.gmcs,evans:2014.sfe, heyer:2016.clumps}. In this case the measured mass traces the stars formed over the characteristic evolutionary timescale $\tau_{SF}$ for the class of YSO that is being counted, typically taken to be $\unit[0.5]{Myr}$ and $\unit[2]{Myr}$ for Class I and II YSOs respectively \citep{evans:2009.sfe}. 

Whatever the stellar mass tracer, the characteristic lifetime $\tau_{tr}$ introduces certain biases in the inferred stellar mass. If the star-forming lifetime of a cloud scales with the freefall time, then $M_{\star,tr}/M_{\star} \sim {\tau_{tr}}{t_{ff}}$, so $M_{\star,tr}$ would typically underestimate $M_{\star}$ in less-dense clouds that have longer freefall times. On the other hand, assuming $\dot{M}_\star = M_{\star,tr}/\tau_{tr}$ provides a reasonably accurate estimate of the SFR of clouds with lifetimes longer than $\tau_{tr}$, averaged over $\tau_{tr}$, but if the star formation has only been occurring for a shorter time $\Delta t \ll \tau_{tr}$, this method will under-estimate the true SFR by a factor $\sim \Delta t/\tau_{\rm tr}$.

When comparing simulations to observations, we will model stellar mass tracers from simulation data in a straightforward manner, simply taking $M_{\star,tr}$ to be the mass of the star particles that formed more recently than $\tau_{tr}$.

\subsection{Gas Mass Tracers}

The ro-vibrational lines of molecular hydrogen are not excited in cold molecular clouds, so it is also necessary to use a tracer to measure $M_{mol}$. Most commonly, this tracer is CO, the second most common molecule in the cold ISM and its primary coolant. $M_{mol}$ can be estimated by measuring the total luminosity $L_{CO}$ of a CO rotational transition and convert this to a gas mass via the CO-to-${\rm H}_2$ conversion factor $X_{CO}$ \citep{bolatto:2013.xco}. Most studies of molecular clouds measure the brightest line, the $^{12}{\rm CO}$ $J=1\rightarrow 0$ transition, which traces the gas mass of molecular number density $n_{{\rm H}_2} > 100\,{\rm cm}^{-3}$. Higher transitions, or emission from the less-abundant $^{13}{\rm CO}$ species, trace higher densities, $>10^3\,{\rm cm^{-3}}$.

In addition to CO observations, we shall consider observations of dense clumps within GMCs. Dense clumps are typically traced by the ${\rm HCN}\,J=1\rightarrow 0$ transition, which generally have been understood to trace the gas mass of density $10^4\,{\rm cm^{-3}}$ or greater \citep{gao.solomon:2004a.hcn}, however recent work has brought into question whether HCN emission really originates from such high density gas \citep{kauffmann:2017.hcn,goldsmith:2017.electron.excitation}. Gas at this density is believed to have a more direct relationship with star formation, as there is a proportional relationship between FIR luminosity and HCN luminosity on the scale of galaxies \citep{gao.solomon:2004a.hcn, wu:2005.clumps} and dense clumps within the Milky Way \citep{wu:2010.clumps}. This is also roughly the threshold above which \citet{lada:2010.gmcs} proposed a linear relationship between gas mass (derived from dust extinction mapping) and the SFR derived from YSO counts. However, the correspondance between the gas mass actually measured by \citet{lada:2010.gmcs} and the gas mass denser than $10^4\mathrm{cm}^{-3}$ is questionable \citep{clark.glover:2014}. We will be able to examine this relationship in our simulations (\S \ref{section:densegas}).

All of these methods of tracing the gas distribution of a cloud have their own uncertainties, biases, and limitations. A fully rigorous comparison between theory and observations would use a calculation that models the relevant dust and molecular abundances self-consistently. One would then model the species' observed emission or extinction self-consistently via a radiative transfer calculation to produce mock observations to which the same analysis can be applied as the actual observation. We consider this to be beyond the scope of this work, and throughout we shall simply compare the simulated gas mass directly with observationally-reported gas masses. We will model $M_{mol}$ from the simulation data in a straightforward manner, using the \citet{krumholz:2011.molecular.prescription} prescription to derive the abundance of ${\rm H}_2$, and assuming that a faithful tracer of ${\rm H}_2$ is available.

\subsection{Data Compilation}
\label{sec:compilation}
We will compare simulations with measurements from several recent studies that have measured $\epsilon$ and $\epsilon_{ff}$ in both GMCs and dense clumps in the Milky Way with various methodologies. These studies, their tracers, and their results are summarized in Table \ref{table:observations}. We consider star-forming GMC data from \citet{evans:2014.sfe}, \citet{vuti:2016.gmcs}, and \citet{lee:2016.gmc.eff}. We also consider data from star-forming dense clumps from \citet{wu:2010.clumps} and \citet{heyer:2016.clumps}.

All of the studies in Table \ref{table:observations} report values for $\epsilon_{ff,obs}$, but only \citet{lee:2016.gmc.eff} reported $\epsilon_{obs}$ (their $\epsilon_{br}$). Where stellar masses are not reported, the measured stellar mass $M_{\star,tr}$ is obtained by simply multiplying reported SFRs by the tracer lifetime $\tau_{tr}$. For \citet{vuti:2016.gmcs}, we compute it by multiplying the SFRs inferred from the clouds' $\unit[24]{\mu m}$ luminosities by the ionization-weighted mean stellar lifetime $\tau_{MS}=\unit[3.9]{Myr}$. We compute the stellar mass in the \citet{wu:2010.clumps} clumps by assuming that their IR luminosities are due to reprocessed radiation from a stellar population with a \citet{kroupa:imf} IMF, and hence:
\begin{equation}
M_{\star,tr} = 8 \times 10^{-4} \left(\frac{L_{IR}}{L_\odot}\right) M_\odot
\end{equation}
and we consider only those clumps with $L_{IR}>\unit[10^{4.5}]{L_\odot}$ so that the IMF is well-sampled \citep{heiderman:2010.gmcs}. The stellar mass obtained directly from YSO counting by \citet{heyer:2016.clumps} is a lower bound on $M_{young}$, while an upper bound is obtained by correcting the total stellar mass assuming a Kroupa IMF. Throughout, we will take the IMF-corrected SFEs computed in this way, but emphasize that these are in fact upper bounds. For consistency with \citet{heyer:2016.clumps}, we estimate SFRs in the \citet{wu:2010.clumps} clumps via Equation \ref{eq:effobs} assuming the same $\tau_{tr}=0.5{\rm Myr}$ as for \citet{heyer:2016.clumps}. Under the above assumptions, SFEs for both catalogues of dense clumps are in good agreement.

\begin{table*}
\begin{tabular}{l|l|l|l|l|l|l|l|l|l|l|l|l}
$M$ ($\mathrm{\msun}$) & $R$ ($\mathrm{pc}$) & Random Seed & Modifications & $t_{ff}$ ($\mathrm{Myr}$) & $t_{2\sigma}$ ($\mathrm{Myr}$) & $\log \frac{n_{{\rm H}_2}}{{\rm cm}^{-3}}$ & $\log \epsilon_{int}$  & $\log \epsilon_{obs}$ &  $\log {\epsilon}_{ff,obs} $ \\ 
\hline
$2\times 10^4$ &   10 & 1 &  & 3.79 & 2.48 & 1.83 & -1.39 & ${-0.79}_{-1.08}^{-0.68}$ & ${-0.86}_{-1.45}^{-0.52}$\vspace{0.1cm} \\
$2\times 10^4$ &   10 & 1 & No Feedback & 3.79 & 7.27 & 1.83 & -0.23 & ${-0.18}_{-1.08}^{-0.07}$ & ${-0.40}_{-1.49}^{0.12}$\vspace{0.1cm} \\
$2\times 10^4$ &   10 & 2 &  & 3.79 & 3.08 & 1.83 & -1.37 & ${-1.28}_{-1.69}^{-0.96}$ & ${-1.44}_{-2.10}^{-0.70}$\vspace{0.1cm} \\
$2\times 10^4$ &   10 & 3 &  & 3.79 & 3.19 & 1.83 & -1.36 & ${-1.06}_{-1.40}^{-0.02}$ & ${-1.06}_{-1.86}^{0.20}$\vspace{0.1cm} \\
$2\times 10^5$ &   30 & 1 &  & 6.22 & 5.12 & 1.39 & -1.43 & ${-1.45}_{-2.44}^{-0.91}$ & ${-1.58}_{-2.67}^{-0.64}$\vspace{0.1cm} \\
$2\times 10^5$ &   30 & 1 & No Feedback & 6.22 & 14.19 & 1.39 & -0.16 & ${-0.12}_{-1.20}^{-0.04}$ & ${0.16}_{-1.39}^{0.43}$\vspace{0.1cm} \\
$2\times 10^5$ &   30 & 2 &  & 6.22 & 6.64 & 1.39 & -1.45 & ${-1.48}_{-2.10}^{-0.99}$ & ${-1.57}_{-2.32}^{-0.63}$\vspace{0.1cm} \\
$2\times 10^5$ &   30 & 3 &  & 6.22 & 5.06 & 1.39 & -1.42 & ${-1.35}_{-2.28}^{-0.99}$ & ${-1.47}_{-2.53}^{-0.67}$\vspace{0.1cm} \\
$2\times 10^6$ &  100 & 1 &  & 11.98 & 15.83 & 0.825 & -1.45 & ${-1.55}_{-2.64}^{-0.99}$ & ${-1.38}_{-2.60}^{-0.28}$\vspace{0.1cm} \\
$2\times 10^6$ &  100 & 1 & No Feedback & 11.98 & 28.61 & 0.825 & -0.14 & ${-0.44}_{-1.62}^{-0.09}$ & ${-0.21}_{-1.54}^{0.42}$\vspace{0.1cm} \\
$2\times 10^6$ &  100 & 1 & No Radiative Feedback & 11.98 & 13.72 & 0.825 & -1.10 & ${-1.33}_{-2.21}^{-0.54}$ & ${-1.17}_{-2.16}^{0.01}$\vspace{0.1cm} \\
$2\times 10^6$ &  100 & 1 & No SNe & 11.98 & 14.19 & 0.825 & -1.46 & ${-1.57}_{-2.43}^{-1.03}$ & ${-1.41}_{-2.39}^{-0.46}$\vspace{0.1cm} \\
$2\times 10^6$ &  100 & 1 & No OB Winds & 11.98 & 12.66 & 0.825 & -1.59 & ${-1.52}_{-2.36}^{-1.27}$ & ${-1.36}_{-2.31}^{-0.75}$\vspace{0.1cm} \\
$2\times 10^6$ &  100 & 1 & Radiative Feedback Only & 11.98 & 14.89 & 0.825 & -1.59 & ${-1.39}_{-2.26}^{-1.00}$ & ${-1.20}_{-2.22}^{-0.37}$\vspace{0.1cm} \\
$2\times 10^6$ &  100 & 1 & SN Feedback Only & 11.98 & 12.43 & 0.825 & -0.96 & ${-1.03}_{-2.22}^{-0.65}$ & ${-0.90}_{-2.18}^{-0.15}$\vspace{0.1cm} \\
$2\times 10^6$ &  100 & 2 &  & 11.98 & 15.83 & 0.825 & -1.53 & ${-1.65}_{-2.77}^{-1.22}$ & ${-1.51}_{-2.71}^{-0.63}$\vspace{0.1cm} \\
$2\times 10^6$ &  100 & 3 &  & 11.98 & 10.79 & 0.825 & -1.57 & ${-1.70}_{-3.00}^{-1.09}$ & ${-1.62}_{-2.97}^{-0.60}$\vspace{0.1cm} \\

\end{tabular}
\caption{Parameters and basic results of the simulations. (1) $M$: the initial total gass mass. (2) $R$: the initial cloud radius. (3) Random seed: the random realization of isothermal MHD turbulence used for the initial conditions. (4) Modifications: variations of subset of feedback mechanisms included. (5) $t_{ff}$: the initial freefall time computed from the mean density $3M/4\pi R^3$. (6) $t_{2\sigma}$: the length of time between the $\pm 2\sigma$ star formation times. (7) $n_{{\rm H}_2}$: volume-averaged initial number density of ${\rm H}_2$. (8) $\epsilon_{int}$: fraction of $M$ converted to stars by the end of the simulation. (9-11) Star formation efficiencies, see Table \ref{table:sfe.review}.  $\log \epsilon_{obs}$ and $\log \epsilon_{ff,obs}$ are given in the format ${\rm median}^{+\sigma}_{-\sigma}$, with the quantiles computed over ``observable'' lifetime of the cloud, during which massive stars are present and the cloud is above the assumed molecular gas surface density sensitivity threshold of $\unit[10]{\msun\,pc^{-2}}$ (see \S\ref{section:sfeobs.methods}).}
\label{table:simresults}
\end{table*}

\section{Simulations}
We perform a suite of 3D MHD simulations of GMC collapse, star formation, and cloud disruption with the {\small GIZMO}\footnote{A public version of this code is available at \href{http://www.tapir.caltech.edu/~phopkins/Site/GIZMO.html}{\url{http://www.tapir.caltech.edu/~phopkins/Site/GIZMO.html}}.} code \citep{hopkins:gizmo,hopkins:gizmo.mhd}, with the prescriptions for star formation, cooling and stellar feedback developed for the Feedback In Realistic Environments (FIRE) simulations \citep{hopkins:2013.fire}\footnote{\url{http://fire.northwestern.edu}}. The simulations are similar in methodology to the \citet{grudic:2016.sfe} simulations, using the Meshless Finite Mass (MFM) Lagrangian MHD method \citep{hopkins:gizmo.mhd} and feedback prescriptions as in FIRE-2 \citep{fire2}. However, the simulations differ in the following ways that we shall describe in turn: the initial conditions (ICs), the star formation prescription, and a modification of the feedback routines that accounts for the effects of under-sampling the IMF when the total stellar mass is small. The reader is referred to \citetalias{grudic:2016.sfe} paper for a description of the general results and behaviour of this type of simulation, and to \citet{fire2} for the details of the numerical implementations of feedback and ISM physics.

\subsection{Initial Conditions}
To model Milky Way GMCs more closely than \citetalias{grudic:2016.sfe}, we simulate three points in mass-radius parameter space, with masses $M=2\times \unit[10^4]{\msun}$, $\unit[2\times 10^5]{\msun}$, and $\unit[2\times 10^6]{\msun}$, and radii $R=\unit[10]{pc}$, $\unit[30]{pc}$, and $\unit[100]{pc}$ respectively, for a mean surface density of $64\,M_\odot{\rm pc}^{-2}$. This parameter space
falls within the range of parameters in which most star-forming GMCs in the MWG lie \citep[e.g.][]{miville:2017.gmcs}. However, we emphasize that selecting a single surface density is not fully representative of a real GMC population, and since we expect the SFE to be dependent on the surface density, we expect there to be residual variations in SFE that this parameter study does not account for.

Unlike \citetalias{grudic:2016.sfe}, the initial velocity field has no bulk rotation component to support it at constant mean surface density -- it is dominated by turbulent motions, which is more consistent with the properties of GMCs in the Local Group, which have quite weak rotation \citep{braine:2018.gmcs}. \citet{krumholz:2012.orion.rhd} found that the initial rise of the SFR was artificially fast when an initial tophat-density distribution was used, as in \citetalias{grudic:2016.sfe}. As this can potentially affect the SFE observables of interest, we follow \citet{krumholz:2012.orion.rhd} by using ICs extracted from a simulation of driven isothermal supersonic MHD turbulence without self-gravity. The turbulent forcing is realized as a Orstein-Uhlenbeck process as in \citet{bauer.springel:2012}, with purely solenoidal forcing, normalized so that the RMS Mach number saturates to $\sim 10$ and the turbulent plasma $\beta$ to $\sim 20$ \citep[e.g.][]{federrath:supersonic.turb.dynamo}. For each set of cloud parameters, we effectively sample three independent statistical realizations of the turbulent ICs by extracting snapshots separated by 10 crossing times each. From each of these snapshots, we excise a sphere centred upon the density-weighted centre of mass and rescale the particle masses, positions, and velocities to achieve the desired mass and radius and virial parameter $\alpha_{vir}=\frac{2 E_{turb}}{|E_{grav}|}=2$. The magnetic field is rescaled to preserve the turbulent plasma $\beta$ from the original turbulent box simulation. In all simulations, the gas mass is initially resolved in $10^6$ particles. 

\subsection{Star Formation}
We handle star formation with an accreting sink particle prescription, derived from \citet{Federrath_2010_sink_particle} and described fully in \citet{guszejnov:2018.isothermal}. To summarize, gas cells are converted to sink particles when they are self-gravitating at the resolution scale (including thermal, turbulent, and magnetic energy contributions), or equivalently, the effective Jeans mass is no longer resolvable. They must also be a local density maximum within their hydrodynamic stencil of nearest neighbor cells, and be a site of converging flow ($\nabla \cdot {\bf v} < 0$). Gas cells are accreted by an existing sink particle if they are gravitationally bound to it and fall within an accretion radius of $0.01{\rm pc}$. This is a more appropriate method for this problem than the discretization of stellar mass into equal-mass star particles used in \citetalias{grudic:2016.sfe}, as the mass resolution of the simulations is always $\ll 100M_\odot$, sufficient to resolve the formation of some stellar-mass objects.

\subsection{Stellar Feedback}
A fully self-consistent determination of the stellar feedback budget would require a self-consistent treatment of massive star formation, which is currently an open physics problem that is sensitive to radiative transfer physics on scales of $1000\,{\rm AU}$ or less \citep{zinnecker:2007.massive.sf,krumholz:2009.massive.sf,tan:2014.massive.sf,rosen:2016.massive.sf}, which we do not resolve in our simulations. Therefore, in the spirit of \citetalias{grudic:2016.sfe}, we choose not to attempt to model the IMF self-consistently, and assume that the mass, energy, and momentum fluxes from stellar feedback are consistent with that from a single stellar population with a well-sampled \citet{kroupa:imf} IMF. We discretize this feedback budget among sink particles as in \citet{sormani:2016.imf.sampling} and \citet{su:2017.discreteness}, giving each particle a discrete number of `O-stars' sampled from a Poisson distribution with mean $\lambda = \frac{\Delta m}{100\msun}$, where $\Delta m$ is the mass of the particle. The O-star number is incremented by a Poisson-sampled value whenever a sink particle is spawned or accretes. This captures some of the effect of under-sampling the IMF in low-mass clusters. Fluxes due to stellar feedback are scaled in proportion to the number of O stars a particle has, and in the limit where the total stellar mass is $\gg 100 \msun$, the IMF-averaged stellar feedback budget is recovered.

\section{GMC Star Formation Histories}
\begin{figure*}
\includegraphics[width=\textwidth]{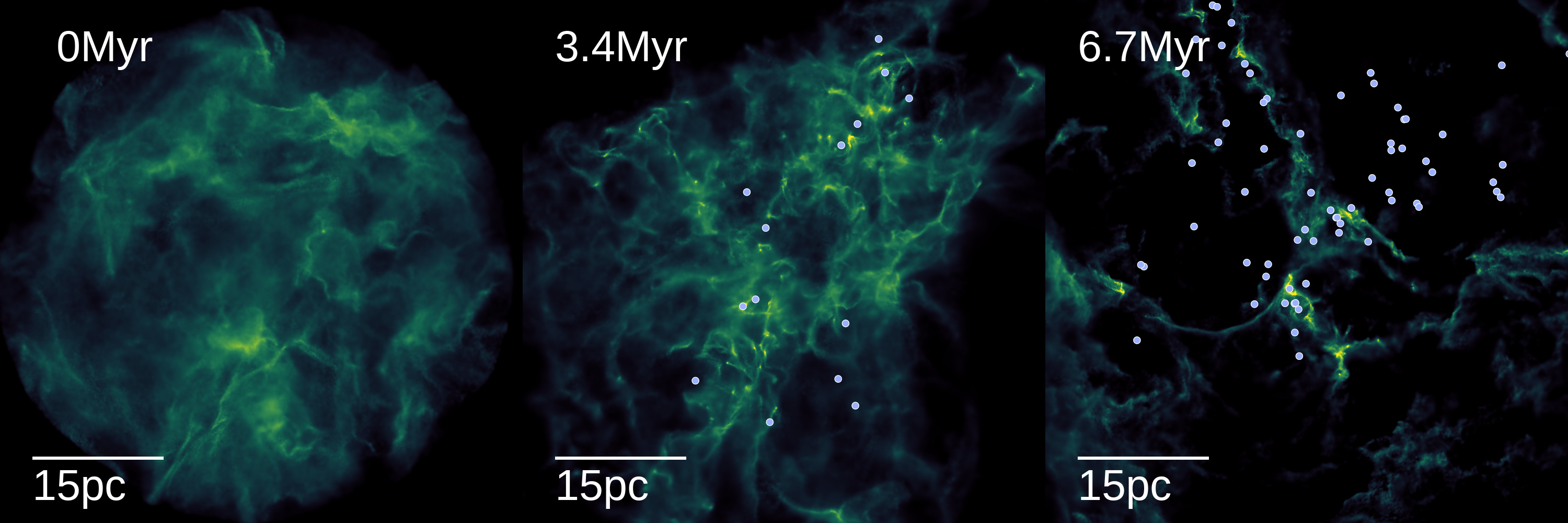}
\caption{Gas surface density map showing the evolution of a simulated GMC with mass $2\times 10^5 \msun$ and radius $30\mathrm{pc}$, with a dynamic range of $10-1000\msun\,{\rm pc}^{-2}$. The cloud undergoes a turbulent, disordered collapse into stars until the combined feedback of massive stars (shown as dots) destroys the cloud. A movie of this sequence can be found at \url{http://www.tapir.caltech.edu/~mgrudich/M2e5_R30.mp4}.}
\label{fig:render}
\end{figure*}


\subsection{True SFE  Values}
\begin{figure*}
\includegraphics[width=\columnwidth]{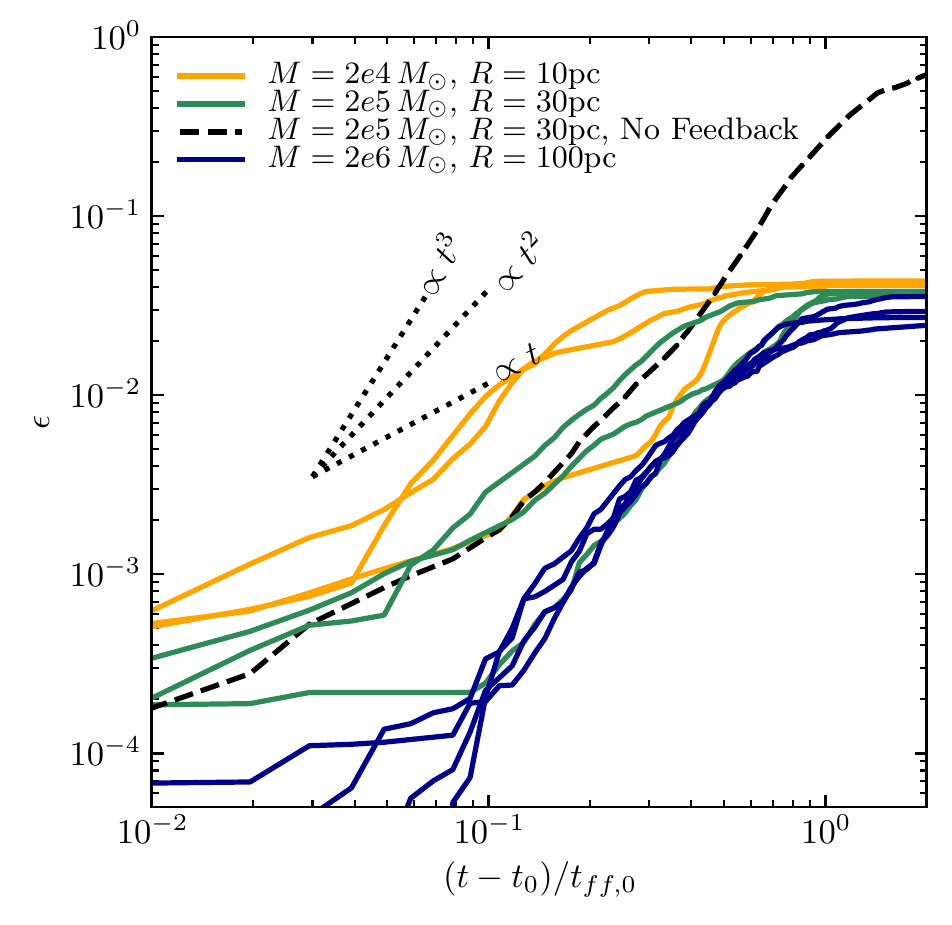} \includegraphics[width=\columnwidth]{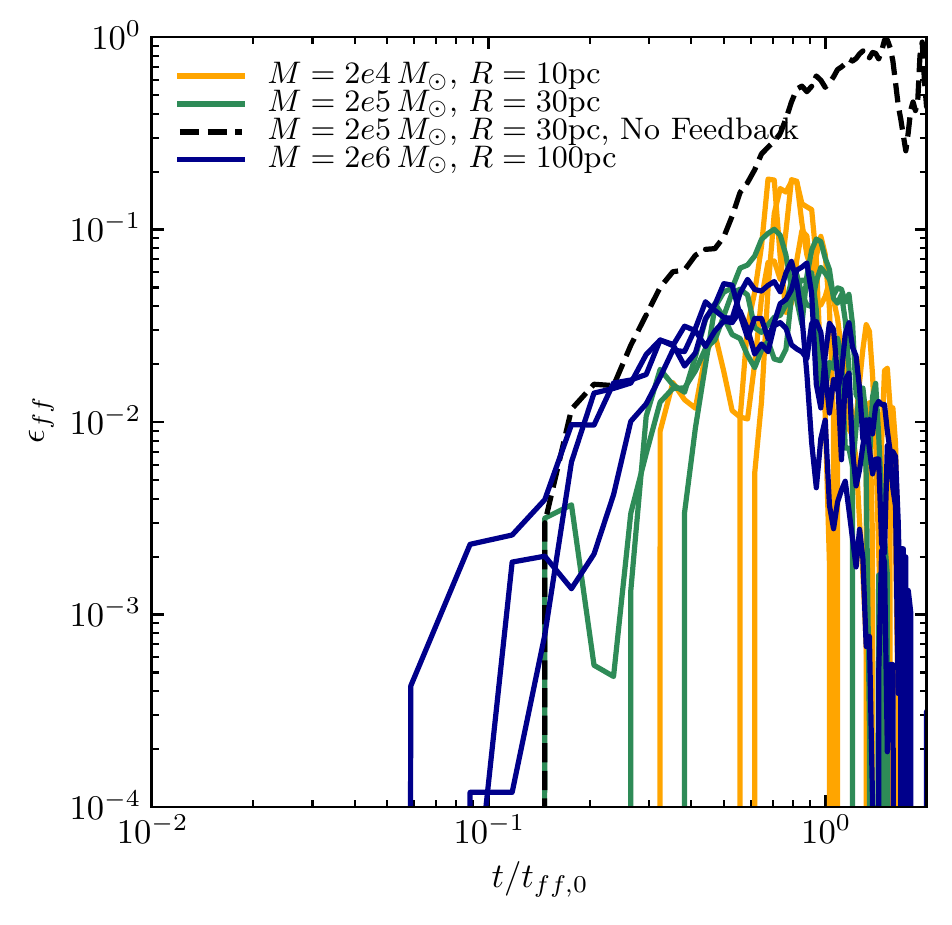}
\caption{Star formation histories of the simulations. {\it Left}: the fraction of the initial gas mass converted to stars $M_\star/M_{tot}$ as a function of time since the formation of the first sink particle at $t_0$, in units of the global cloud freefall time $t_{ff}$. In most cases, the stellar mass is $\propto (t-t_0)^2$ as the SFR ramps up, so there is intrinsic variation in the SFR that would translate into a range of observed $\epsilon_{ff}$. Eventually, sufficient stellar mass forms to disrupt the cloud via stellar feedback, and $M_\star/M \rightarrow \epsilon_{int}$, which is on the order of a few per cent when stellar feedback is included, but approaches $100\%$ in absence of feedback.}
\label{fig:tff_sfe}
\end{figure*}
In all simulations, the cloud initially collapses in a disordered manner, with multiple centres of collapse and little initial global contraction (see Figure \ref{fig:render}). Stars eventually form in dense, gravitationally-bound subregions until stellar feedback is sufficient to halt collapse and accretion locally.  Eventually, a sufficient stellar mass forms that the entire cloud is disrupted by stellar feedback and star formation ceases. At this point, a fraction $\epsilon_{int}$ of the initial gas mass has been converted to stars. As has generally been found in similar simulations, an order-unity fraction of the gas mass is rapidly converted to stars on the freefall timescale when stellar feedback is neglected. When stellar feedback is included, $\epsilon_{int}$ is always a few per cent and does not vary greatly across our parameter space. This is because all of the cloud models have the same mean initial surface density, which determines $\epsilon_{int}$ for feedback-disrupted self-gravitating molecular clouds \citep{fall:2010.sf.eff.vs.surfacedensity,grudic:2016.sfe}. This is in good agreement with the median value of $\epsilon_{obs}$ found in Milky Way GMCs (see Table \ref{table:observations}), however much greater and smaller values are also observed, which we will address in \S \ref{section:sfeobs}. The variation in $\epsilon_{int}$ for different random realizations of a given point in parameter space is also quite small ($<\unit[0.1]{dex}$). Therefore, even when fully-turbulent initial conditions are considered, the instrinsic SFE variations due to variations in specific microstates of the initial conditions clearly cannot explain the observed range of SFE values.

In Table \ref{table:simresults}, we report $T_{2\sigma}$, the length of time containing $95\%$ of star formation, as well as $\epsilon_{int}$ and the average value of $\epsilon_{ff}$ over entire the star formation history. Here the freefall time used to compute $\epsilon_{ff}$ is that computed from the initial volume-averaged density, $t_{ff,0} = \frac{\pi}{2\sqrt{2}}\sqrt{\frac{R^3}{GM}}$, however in \S \ref{section:sfeobs} we will consider the effects of a dynamic mean cloud density upon the observed $\epsilon_{ff}$. The volume-averaged density tends to increase slightly due to turbulent dissipation in the initial stages of cloud collapse, but it then decreases rapidly as stellar feedback launches outflows.


In general $t_{2\sigma}\sim t_{ff,0}$, so most star formation is found to take place within a single initial freefall time (although this can be several {\em local} free-fall times in the denser gas that forms as fragmentation proceeds), as found in \citetalias{grudic:2016.sfe} and similar works. In all instances, the SFR, and hence $\epsilon_{ff}$, is found to vary significantly throughout the GMC lifetime. The SFR tends to continue to increase until star formation is quenched abruptly when the molecular cloud is disrupted by feedback. In Figure \ref{fig:tff_sfe} we present the detailed star formation histories of all simulations, plotting $\epsilon$ and $\epsilon_{ff}$ as functions of time in panels 1 and 2 respectively. We find that the initial growth of $\epsilon$ from the beginning of star formation is superlinear, with the exception of the $M=\unit[2\times 10^4]{\msun}$ runs, and is typically initially well-described by a power-law with index close to $2$, as has been predicted analytically and found in hydrodynamic simulations without stellar feedback \citep{murray.chang:2015,lee:2015.gravoturbulence,murray:2017} and with a more limited subset of feedback channels \citep{raskutti:2016.gmcs,vazquez:2015,geen:2017}. However, this state of affairs does not continue indefinitely, and stellar feedback eventually causes the SFR to level off and eventually fall to 0. 

The lowest-mass cloud models, with $M=2\times 10^4\msun$, have the noisiest star formation histories (e.g. Fig \ref{fig:tff_sfe} panel 2). These clouds only ever only form $4-8$ massive stars before being disrupted, so the proportional effect of an individual massive star on the overall cloud evolution is much greater than in the more massive clouds, which form tens to hundreds of massive stars, making the onset of feedback effectively ``smoother''. The more pronounced effect of the discreteness of massive stars also explains the shallower, nearly linear initial growth of the 
SFR in the low-mass clouds compared to the more massive ones. We have confirmed that when feedback is disabled, the growth in stellar mass is superlinear as in the more massive cloud models.

Because $t_{2\sigma} \sim t_{ff,0}$, $\epsilon_{ff}$ is on average of the same order as $\epsilon_{int}$, which is $\sim 100\%$ without stellar feedback and several per cent with stellar feedback. To summarize, we find the key results of the simulations of \citetalias{grudic:2016.sfe} concerning the true star formation efficiencies of molecular clouds still hold for the more realistic GMC and stellar feedback models we have considered here. Most star formation occurs within a single $t_{ff,0}$, and during this time only several per cent of the initial gas mass is converted to stars, because this fraction is sufficient to disrupt the cloud via stellar feedback. This fraction $\epsilon_{int}$ is approximately the same for clouds of the same surface density, and depends upon spatial scale only weakly. We also find a time-varying $\epsilon_{ff}$ that initially grows in a manner similar to what was found in previous calculations that did not include stellar feedback \citep{murray.chang:2015, lee:2015.gravoturbulence, murray:2017}, however stellar feedback eventually halts the growth. We will now consider how these highly dynamic star formation histories would imprint upon the observed distributions of star formation efficiencies.



\subsection{Tracer-Inferred Values}
\label{section:sfeobs}
\begin{figure*}
\includegraphics[width=\columnwidth]{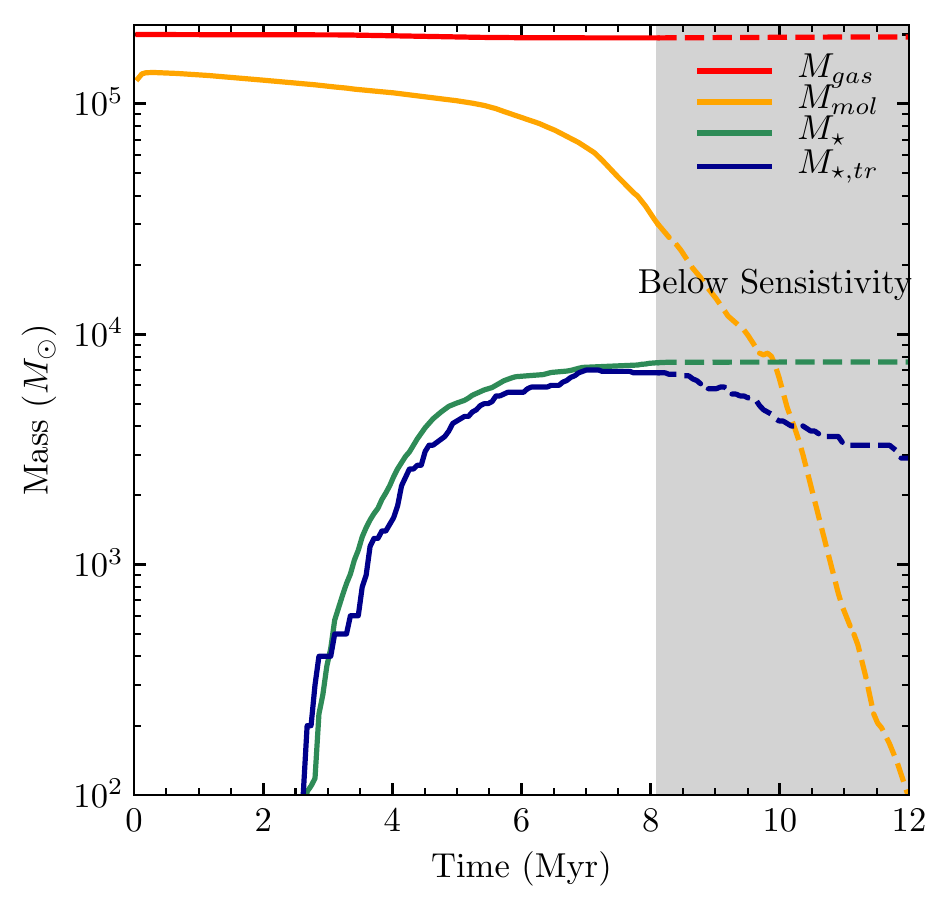}
\includegraphics[width=\columnwidth]{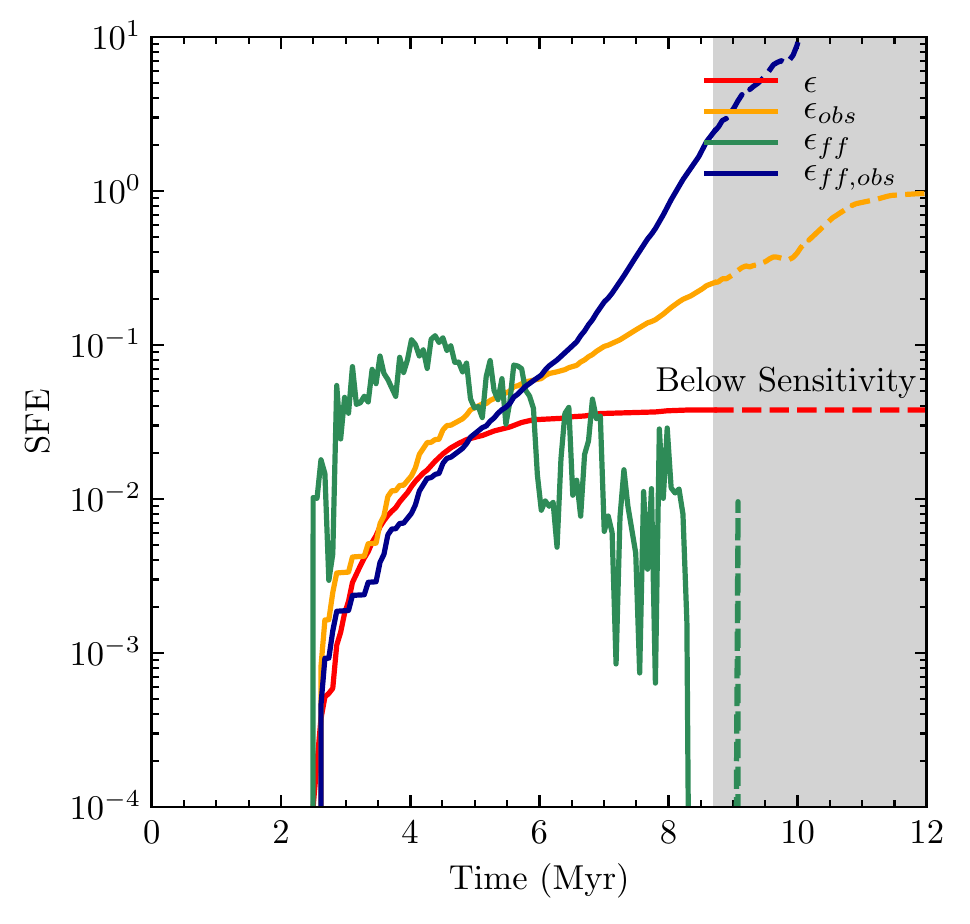}
\caption{Comparison of the true and observable masses and SFEs in the simulation with $M=2\times 10^5 M_\odot$, $R=30{\rm pc}$ and random seed 1. Solid lines denote the portion of the lifetime during which the cloud would plausibly be identified in a star-forming GMC catalogue according to the criteria given in \S\ref{section:sfeobs.methods}. {\it Left:} total gas mass, molecular gas mass, total stellar mass, and traced stellar mass (with $\tau_{tr}=3.9{\rm Myr}$) as a function of time. Both the observed molecular gas and stellar mass underestimate the true gas and stellar masses present at all times. The observed molecular gas mass decreases rapidly once cloud disruption begins. The observed stellar mass also decays to 0 at late times, but not necessarily as rapidly as the observed gas mass, possibly leading to large observed SFE. {\it Right:} various measures of SFE (see \S \ref{section:sfe.review} and Table \ref{table:sfe.review})}
\label{fig:mi}
\end{figure*}
\begin{figure*}
\includegraphics[width=\textwidth]{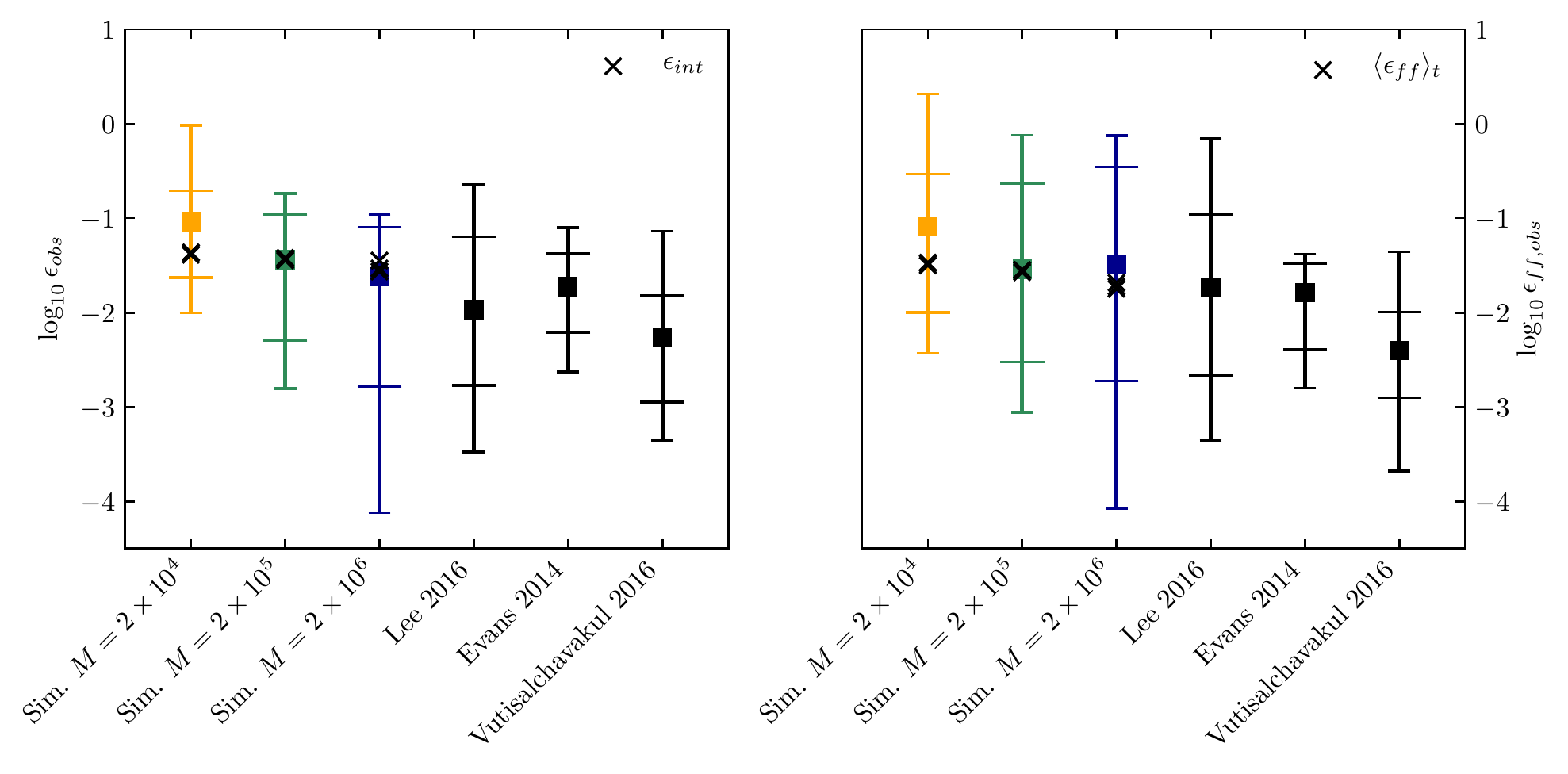}
\caption{Quantiles of the distributions of $\epsilon_{obs}$ (left) and $\epsilon_{ff,obs}$ (right) arising from ``observing" simulated molecular clouds at random points in their observable star formation history, compared with observed cloud SFEs in the Milky Way. Points plot median values, inner whiskers plot $\pm\sigma$ quantiles, and outer whiskers plot $\pm 2\sigma$ quantiles. Both $\epsilon_{obs}$ and $\epsilon_{ff,obs}$ can both be measured to be considerably higher or lower than the true typical efficiencies (overlaid as black markers), depending on when they are measured. The spread in $\epsilon_{obs}$ is as great as $0.8{\rm dex}$ and the spread in $\epsilon_{ff,obs}/\langle \epsilon_{ff}\rangle$ is $0.8-1.1 {\rm dex}$. The median observed $\epsilon_{obs}$ is anticorrelated with cloud mass, as has been observed \citep{murray:2010.sfe.mw.gmc,lee:2016.gmc.eff}. Details of how $\epsilon_{obs}$ and $\epsilon_{ff,obs}$ are modeled are described in \S\ref{section:sfeobs.methods}.}
\label{fig:deltaquantiles}
\end{figure*}
\subsubsection{Modeling of observables}\label{section:sfeobs.methods}
To forward-model $\epsilon_{obs}$ and $\epsilon_{ff,obs}$ from the simulations, we must estimate the observationally-inferred $M_\star$ and $M_{mol}$. One possibility is to perform ISM chemistry and radiative transfer calculations to directly model the observed emission maps, and apply the same procedure for identifying clouds and correlating them with young stars as was used in a particular study. This would be necessary to compare with observations in detail. This approach is possible in principle, but here we merely aim to explain various general features and trends from many studies of widely varying methodology, so in this initial investigation we use simple approximations, leaving a more detailed treatment to future work.

We estimate the observed $M_{\star,tr}$ in the simulation data by taking $M_{\star, live}$, approximately the mass in stars younger than the ionization-weighted mean stellar lifetime $3.9\,{\rm Myr}$ \citep{murray:2010.sfe.mw.gmc}. We estimate $M_{mol,tr}$ by assuming that a perfect tracer of ${\rm H_2}$ is available, and simply take the actual molecular gas mass $M_{mol}$. We do this by calculating the molecular fraction $f_{H_2}$ of each gas cell according to the analytic prescription of \citet{krumholz:2011.molecular.prescription}, which was found to agree well with detailed chemistry and radiation transfer calculations. This prescription requires an estimate of the effective dust optical depth $\tau_c$ at the position of each gas cell. This is calculated on-the-fly in the simulations using a local Sobolev-like column density estimator \citep{hopkins:fb.ism.prop}, but we compute $\tau_c$ more accurately in post-processing by ray-tracing the dust opacity field to infinity along 64 rays on an equal-area spherical grid. An optical depth $\tau_i$ is thus obtained for each ray, and the effective optical depth $\tau_c$ is that which gives the spherically-averaged extinction over all angles $\theta$ and $\phi$:
\begin{equation}
\exp \left(-\tau_c\right) = \frac{1}{4\pi}\int \exp\left(-\tau\left(\theta,\phi\right)\right)\,{\rm d}\Omega \approx \langle \exp \left(-\tau_i\right) \rangle_i,
\end{equation}
where $\langle \cdot \rangle_i$ denotes the mean value over all rays.

Modeling $\epsilon_{ff,obs}$ requires a measurement of $t_{ff}$, which depends upon the mean cloud density and hence its effective volume. For this, we use the 3D equivalent of the technique used in \citet{miville:2017.gmcs} for deriving effective cloud volumes from CO emission maps. We take the volume of the ellipsoid with axes given by the the eigenvalues of the 3D $f_{{\rm H}_2}$-weighted covariance matrix of the gas distribution. Given eigenvalues $\lambda_i$, the observed effective volume is taken as $V=4\pi R_{eff}^3/3$, where $R_{eff} = \left( \lambda_1 \lambda_2 \lambda_3 \right)^{1/6}$.

Lastly, we must account for observational selection effects and reject simulation data that clearly would not be identified as a data point in a catalogue of star-forming GMCs. We measure all relevant SFE statistics only during the fraction of the GMC's lifetime during which it could possibly be counted as an association between emission from young stars and molecular gas emission. We do this by including only simulation snapshots satisfying two criteria:
\begin{itemize}
\item Both molecular gas and stars younger than $3.9 {\rm Myr}$ are present. This effectively determines the beginning of the observable time interval.
\item The mean molecular gas surface density $\Sigma_{gas}=M_{mol}/\pi R_{eff}^2$ is $>10 M_\odot \,{\rm pc^{-2}}$, corresponding to the $-2\sigma$ quantile of measured mean surface density of the star-forming GMCs in \citet{lee:2016.gmc.eff}. This approximates the latter boundary of the observable cloud lifetime, since $\Sigma_{gas}\rightarrow 0$ as the cloud is disrupted.
\end{itemize}

\subsubsection{Evolution of observables}
In the first panel of Figure \ref{fig:mi} we plot the evolution of the true and observable masses and SFEs in the simulation with $M=2\times 10^5 M_\odot$, $R=30{\rm pc}$, and random seed 1. $M_{mol}$ is initially close to the actual total gas mass present, missing only the gas mass in the low surface density tail of the log-normal turbulent column density PDF \citep[e.g.][]{thompson:2016.eddington.outflows}, which is not self-shielding. We therefore expect that $M_{mol}$ measured in observations is a reasonably faithful estimate of the total gas mass of molecular clouds that have not yet undergone significant star formation, insofar as the tracer-to-${\rm H}_2$ conversion factor is accurate.

As massive stars form, $M_{mol}$ begins to decrease increasingly rapidly as stellar feedback starts to disrupt the cloud. Two physical effects cause this: gas launched in feedback-driven outflows tends to expand to the point that it is no longer self-shielding to the UV background, and gas near massive stars is ionized, forming HII regions. Once the cloud is fully disrupted and star formation has ceased, $M_{mol}$ decays rapidly toward 0 with a roughly exponential behaviour with an e-folding time of only $\sim 0.5\,{\rm Myr}$.

$M_{\star,tr}$ always underestimates $M_\star$, but it is a reasonably good estimate during the initial ramp-up of the SFR because the total stellar mass formed is dominated by the most recently-formed stars. Toward the end of the star formation history, when the SFR starts to drop, $M_{\star,tr}$ begins to underestimate $M_\star$ more significantly, eventually decaying to $0$ after star formation has ceased. The observable masses in all other simulations follow these same general patterns as the run shown in Figure \ref{fig:mi}. However, we do find that more massive clouds tend to have a longer span of time during which $M_{\star,tr}$ underestimates $M_\star$ noticeably. This is due to their longer star-forming lifetimes compared to the tracer lifetime ($t_{ff}\propto M^{1/4}$ at constant $\Sigma$).

In the second panel of Figure \ref{fig:mi} we plot the evolution of the true and observable SFEs that result from these evolving stellar and gas masses. $\epsilon$ increases monotonically toward $\epsilon_{int}$, but $\epsilon_{obs}$ does not necessarily approach a constant because it is a ratio of two rapidly-changing observed masses. During the cloud disruption phase, $M_{mol}$  $\epsilon_{ff}$ rises with the initial increase in SFR, peaks, and decays to 0 as the cloud is disrupted, but again $\epsilon_{ff,obs}$ continues to increase without bound even more rapidly than $\epsilon_{obs}$. This is due to a combination of effects: the ratio $M_{\star,tr}/M_{mol}$ increases rapidly, 
and
the observed $t_{ff}$ also increases due to the expanding effective cloud volume. The observed $\epsilon_{ff,obs}$ can be doubly boosted by orders of magnitude beyond the true $\epsilon_{ff}$, which never exceeds a few per cent.

\subsubsection{Distributions of $\epsilon_{obs}$ and $\epsilon_{ff,obs}$}
The manner in which $\epsilon_{obs}$ and $\epsilon_{ff,obs}$ vary throughout the cloud lifetime will imprint upon the distributions of values observed for an ensemble of clouds at random points in their lifetimes. In Figure \ref{fig:deltaquantiles} we plot the quantiles of these distributions by cloud mass and compare them with the observations of \citet{lee:2016.gmc.eff}, \citet{evans:2014.sfe}, and \citet{vuti:2016.gmcs}. For each simulation we also overlay $\epsilon_{int}$ for each simulation for comparison with $\epsilon_{obs}$ and the true per-freefall SFE averaged over the star-forming lifetime, $\langle \epsilon_{ff}\rangle_t$, for comparison with $\epsilon_{ff,obs}$. 

The $\epsilon_{obs}$ and $\epsilon_{ff,obs}$ distributions from the simulations are able to reproduce all essential features of the observed ones: they are all fairly broad, with significant scatter about a median value on the order of $1\%$. The spreads in $\epsilon_{obs}$ and $\epsilon_{ff,obs}$ dwarf the true variation in $\epsilon_{int}$ and the time-averaged $\langle\epsilon_{ff}\rangle_t$ from one cloud to another.  Heavy lower tails result from the initial gradual growth of the SFR (Figure \ref{fig:tff_sfe}). Excursions of $\epsilon_{obs}$ in excess of $10\%$ occur due to the rapid depletion of molecular gas, and even heavier upper tails are found for $\epsilon_{ff,obs}$ because this effect is combined with an increase in the observed $t_{ff}$ as the cloud expands. 

\subsubsection{Trends in observed SFE with GMC mass}
The median observed SFEs are generally fairly close to the true SFEs. However, although there is no systematic trend in the true $\epsilon_{int}$ with cloud mass, the median $\epsilon_{obs}$ is anticorrelated with cloud mass, scaling as $\epsilon_{obs} \propto M^{-0.25}$. This trend is also found in observations \citep{murray:2010.sfe.mw.gmc,lee:2016.gmc.eff}, with \citet{lee:2016.gmc.eff} finding  $\epsilon_{obs}\propto M^{-0.31}$, and has several possible explanations. 

The use of free-free emission as a stellar mass tracer introduces a selection bias for star-forming regions that host massive stars. Assuming a universal and stochastically-sampled IMF, the effect of this would be to bias measurements toward star-forming regions hosting a stellar mass greater than a certain threshold mass $M_{\star,min}$, above which the IMF is well-sampled and massive stars are expected to be present. This translates into an effective threshold for $\epsilon_{obs}$, $\epsilon_{min} = M_{\star,min}/M \propto M^{-1}$, which accounts nicely for the trend found in \citet{murray:2010.sfe.mw.gmc}, but is much steeper than the trend found in in \citet{lee:2016.gmc.eff}. \citet{murray:2010.sfe.mw.gmc} also noted that uncertainties in $M_{gas}$ will tend to scatter points along a locus $\epsilon_{obs} \propto M^{-1}$, which again is steeper than the trend that is observed in \citet{lee:2016.gmc.eff}.

In the simulations, we find that the trend is due to the fact that more massive clouds have longer lifetimes, so $M_{\star,tr}$ will tend to underestimate $M_\star$. If the cloud lifetime scales $\propto t_{ff}$, which we find, then we expect a scaling $\epsilon_{obs}\propto t_{ff}^{-1} \propto M^{-1/4}$, hence the $\sim 0.5{\rm dex}$ decrease over $2 {\rm dex}$ in cloud mass. This is much closer to the observed scaling, so we favour this explanation.

\citet{ochsendorf:2017.gmcs} also identified an anticorrelation of $\epsilon_{ff,obs}$ with total mass in star-forming complexes in the \citet{lee:2016.gmc.eff} Milky Way clouds as well as the LMC, of strength ranging from $\propto M^{-0.11}$ to $\propto M^{-0.49}$ respectively. Our simulations do not predict a correlation as strong as is observed in the LMC, and thus are not able to explain this trend. As \citet{ochsendorf:2017.gmcs} argued, the greater importance of diffuse, CO-dark gas in the overall mass budget in the lower-metallicity environment of the LMC might explain the difference in the strength of the trend between the Milky Way and the LMC. Since we have assumed that all molecular gas is being traced, this type of effect is not captured in our analysis.

\subsubsection{Conclusions}

We find the spread in SFE in a given sample of clouds
to be driven by two effects:
1) the slow initial growth of stellar populations and the late dispersal of clouds; 
and
2) systematic trends with cloud mass within the sample. 
Both 
effects 
may
explain why the datasets of \citet{evans:2014.sfe} and \citet{vuti:2016.gmcs} have less spread than \citet{lee:2016.gmc.eff}: the samples both span a much narrower range in $M$ and $\Sigma_{gas}$ (for a summary of cloud properties see Table \ref{table:observations}). A narrower range, and in particular a greater lower bound on $\Sigma_{gas}$ would capture less of the the late cloud disruption stage. A narrower range in mass scale will capture less of the systematic scalings with mass scale that we have found.

The normalization of the observed SFEs---set by the physics of stellar feedback as shown in the simulations---is recovered within factors of $\sim$2--3. An order unity discrepancy can be easily accounted for by the systematic errors expected from models (e.g., uncertainties in massive star formation, stellar evolution, and stellar feedback; see discussion in \citetalias{grudic:2016.sfe}) and from observations (e.g., errors in tracer conversion factors and the identification of gravitationally-bound gas).
\citet{vuti:2016.gmcs} point out that the SFR estimator they used underestimates the total star formation in the Milky Way by a factor of 2--3, which may explain why their SFEs are noticeably lower than the simulations and other observations.

In summary, we find that most of the observed spread in the SFE of molecular clouds can be explained by the variation that occurs during the evolution a single cloud, subject to the effects of feedback from massive stars. There is also a spread due to a systematic trend 
between the observable SFE and the bulk properties of the cloud,
but this does {\it not} imply a trend in the true SFE. Indeed, most of the various SFE observations could be attributed to a population of clouds for which $\epsilon_{int}$ actually varies very little. This is in line with a picture where molecular clouds in the Milky Way form with only small spread in $\Sigma_{gas}$ (e.g., due to the properties of supersonic turbulence: \citealt{larson:gmc.scalings, ballesteros.paredes:2011, hopkins:2012.excursion.set}), and as a result do not vary greatly in $\epsilon_{int}$ because it is a function of surface density due to the scalings of self-gravity and stellar feedback \citep[e.g.][]{fall:2010.sf.eff.vs.surfacedensity,grudic:2016.sfe,kim:2018}.

\subsection{Evolution of $\alpha_{vir}$}
\begin{table}
\begin{tabular}{l|l|l}
$\epsilon_{ff,obs}-\alpha_{vir}$ dataset & Kendall $\tau$ & $p$-value \\ \hline
\citet{lee:2016.gmc.eff} & 0.142 & 0.00359 \\
\citet{vuti:2016.gmcs} & 0.194 & 0.0313 \\
\citet{evans:2014.sfe} & -0.0533 & 0.708 \\
\end{tabular}
\caption{Parameters of the Kendall $\tau$-test for correlation between $\alpha_{vir}$ and $\epsilon_{ff,obs}$ in our GMC data compilation. Both significant correlations ($p<0.05$) are positive ($\tau > 0$), consistent with a monotonic increase of both $\alpha_{vir}$ and $\epsilon_{ff,obs}$ with cloud age.}
\label{table:kendall}
\end{table}
The results of this section suggest an interpretation of the upper tails of the SFE distributions in terms of molecular cloud dispersal: molecular gas is destroyed or ejected at the end of the cloud lifetime due to stellar feedback, causing the $M_{gas}$ term in the denominator of the SFE to become small and the inferred SFE to rise, although the actual SFR is dropping. If molecular clouds are initially gravitationally bound, with $\alpha_{vir}\leq 2$, as we have simulated, then a key prediction of this picture is an increase of $\alpha_{vir}$ from the initial bound state to greater values as the cloud evolves. We have verified that when the simulations reach the threshold of detectability, the clouds have $\alpha_{vir} \sim 10-20$, similar to the maximum value observed. However, \citet{lee:2016.gmc.eff} searched for a correlation between the size of the HII bubble associated with a cloud and its virial parameter, and none was found. Since we find $\epsilon_{ff}$ to be a monotonic and fairly sensitive function of the cloud evolutionary stage (Figure \ref{fig:mi}), we may also test for correlations between $\epsilon_{ff,obs}$ and $\alpha_{vir}$ in our data compilation. The results of the Kendall $\tau$-test for correlations between $\epsilon_{ff,obs}$ and $\alpha_{vir}$ are given in Table \ref{table:kendall}. We find a positive correlation between $\alpha_{vir}$ and $\epsilon_{ff}$ for both the \citet{lee:2016.gmc.eff} and \citet{vuti:2016.gmcs} datasets at $2.9\sigma$ and $2.1\sigma$ significance, respectively.

Nevertheless, these correlations are rather weak, and the lack of correlation with HII region size is still puzzling -- clearly the picture is more complicated than a universal evolution from $\alpha_{vir}=2$ to $> 10$. However, many factors might explain the scatter and weakness of the trend in $\alpha_{vir}$ with cloud evolution, either by introducing scatter or by increasing the expected mean measured value of $\alpha_{vir}$. The empirically-measured $\alpha_{vir}$ is \citep{bertoldi.mckee}:
\begin{equation}
\alpha_{vir} = \frac{5 \sigma_{v}^2 R}{GM},
\end{equation}
where $\sigma_{v}$ is the measured 1D velocity dispersion. This only equals the virial ratio $\frac{2 E_{turb}}{|E_{grav}|}$ in the case of a uniform sphere with a flat internal size-linewidth relation. If GMCs are triaxial, intrinsic scatter is immediately introduced by using only 1D and 2D information for $\sigma_v$ and $R$ respectively. Correcting the size-linewidth relation to that of supersonic turbulence raises the threshold for marginal boundness from 2 to $10/3 \sim 3.3$ \citep{miville:2017.gmcs}. It is also possible that the typical virial parameter at which a GMC starts to form stars is even greater than this threshold: a cloud that is not globally bound may still have bound subregions that can collapse and form stars, so theory predicts the SFR to be a continuously decreasing function of $\alpha_{vir}$, rather than a sharp cutoff \citep[e.g.][]{padoan:2012.sfe}. Furthermore, the traditional concept of gravitational boundness of clouds in terms of $\alpha_{vir}$ neglects the fact that GMCs are in a state of supersonic turbulence, and hence are dissipating kinetic energy on a crossing time \citep{gammie.ostriker:1996.mhd.dissipation}. This could potentially allow them to reach higher virial parameters than an equivalent dissipationless system without dispersing. If a significant fraction of star formation is in less-bound clouds, then the actual variation of $\alpha_{vir}$ throughout the observable cloud lifetime might actually be quite modest, weakening any observed correlation. Lastly, it is possible that a significant fraction of molecular emission from a cloud does not originate in the dynamically-active region that is causally connected to the star formation event and directly affected by feedback. A molecular cloud consisting of a diffuse molecular envelope and a more tightly-bound, star-forming core might not be observed to have a large variation in $\alpha_{vir}$ throughout its observable lifetime.

\subsection{Effects of different feedback mechanisms}
Stellar radiation, winds from OB stars, and supernova explosions all contribute to the disruption of clouds, but which of these feedback mechanisms is the most dominant?
In answering this question, we focus upon the cloud model with mass $2 \times 10^6\msun$ and radius $\unit[100]{pc}$, 
because it provides the most dynamically interesting environment
: the cloud lifetime of $\sim 10\rm{Myr}$ is sufficiently long for Type II SNe from the first massive stars to occur during it, but is not so long that the net fluxes of OB winds and stellar radiation from the stellar population are seriously reduced by the deaths of these massive stars. As such, all mechanisms could potentially be important.

We have re-simulated the $2 \times 10^6\msun$, $\unit[100]{pc}$ cloud model with random seed 1, with several combinations of feedback physics, performing three runs that neglect stellar radiation, OB winds, and SNe respectively, as well as runs where radiation and SNe are the only feedback, with results summarized in Table \ref{table:simresults}. Radiative feedback is clearly the most important: when it is neglected, both the true and observed SFE values increase by $\unit[0.3-0.5]{dex}$. However, if the other feedback mechanisms are neglected, the change in SFE is quite small, $\unit[<0.1]{dex}$. Because the star-forming lifetime is $>\unit[3.5]{Myr}$, we find that SNe alone are able to moderate star formation and disrupt the this cloud model, giving a $\epsilon_{int} \sim 10\%$ compared to $\sim 4\%$ with all feedback mechansisms. However, this would not be the case for the smaller cloud models, which evolve on a shorter freefall timescale.

It should be noted that although we find that radiative feedback alone to be sufficient to set the cloud SFE, the interplay of different feedback mechanisms in concert may have other effects not considered here. For example, although SNe may be subdominant in setting the cloud SFE, they can conceivably enhance the terminal momentum of the cloud once it is disrupted. Such an effect could easily be important in the greater galactic context, where feedback supports the galactic disk against collapse \citep{thompson:rad.pressure,ostriker.shetty:2011,cafg:sf.fb.reg.kslaw,orr:2018.kennicutt.schmidt}.

\section{Star Formation in Dense Gas}
\label{section:densegas}

\begin{figure*}
\includegraphics[width=\textwidth]{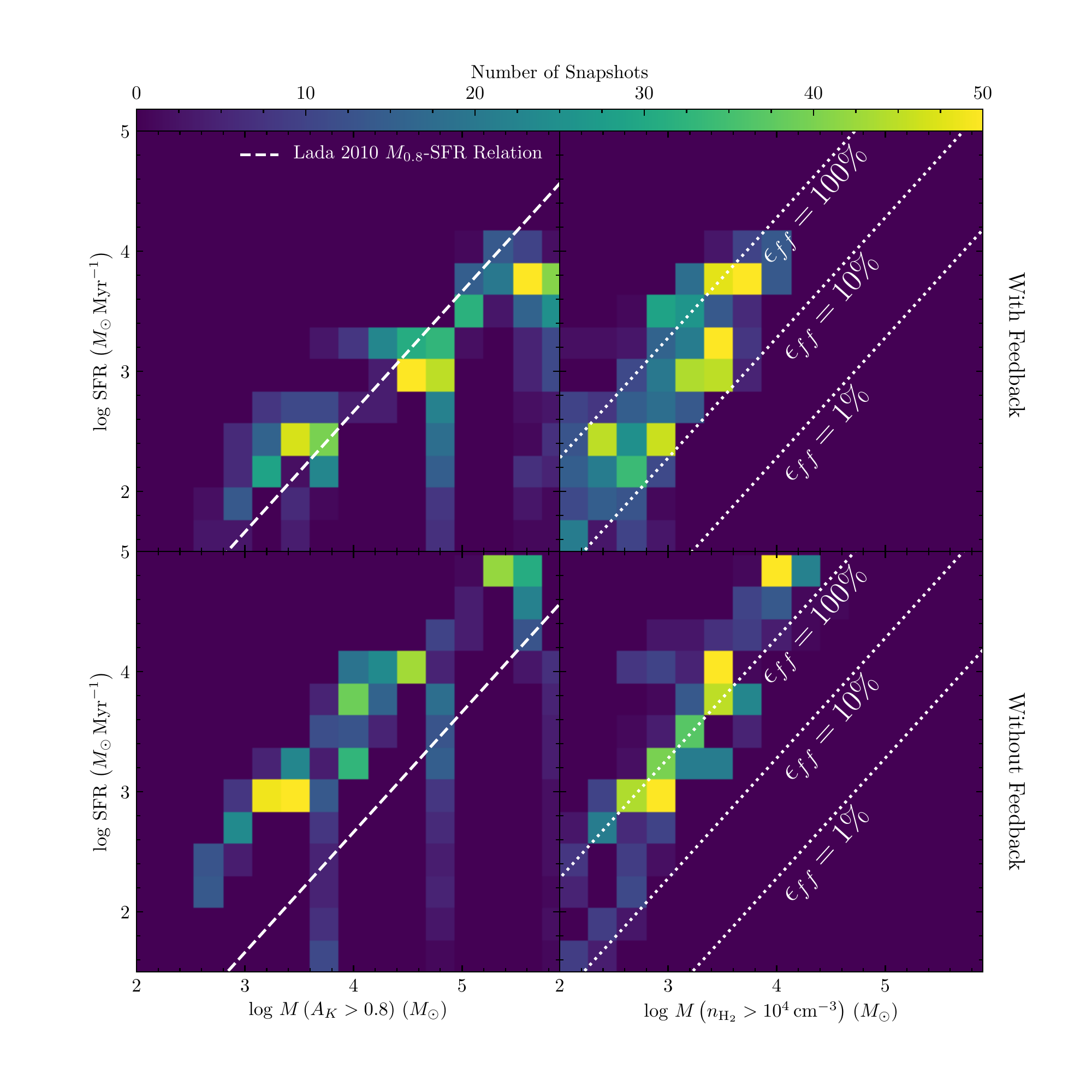}
\vspace{-1.5cm}
\caption{SFR as a function of ``dense'' gas mass in the simulations, for both 2D and 3D density thresholds. We plot 2D histograms of the compilation of all simulation snapshots in $M_{gas}-SFR$ space to give a sense of the relative amount of time spent by the simulations at a given point. {\it Left:} SFR as a function of gas mass above $\unit[0.8]{mag}$ extinction, compared to the \citet{lada:2010.gmcs} relation, for simulations with (top) and without (bottom) stellar feedback. {\it Right:}  SFR as a function of gas mass with molecular gas density $n_{\mathrm{H}_2}$ greater than $\unit[10^4]{cm^{-3}}$, for simulations with (top) and without (bottom) stellar feedback.}
\label{fig:M_v_SFR}
\end{figure*}

\begin{figure}
\includegraphics[width=\columnwidth]{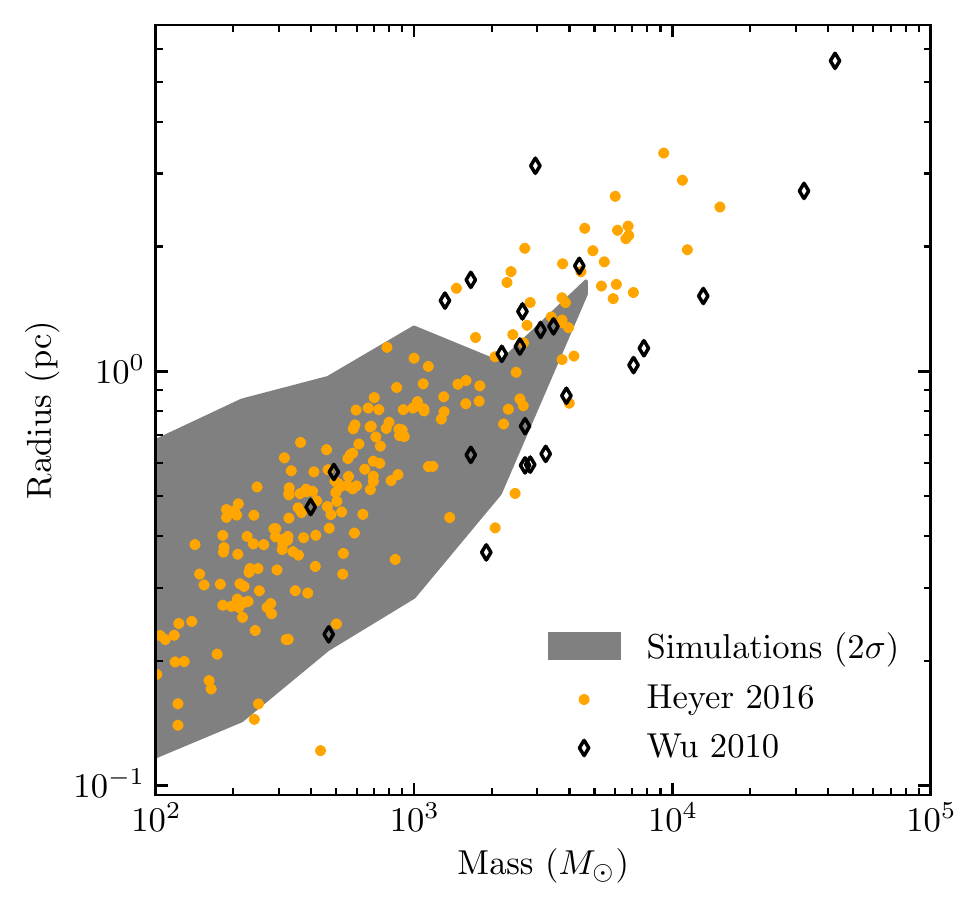}
\caption{Mass versus effective radius for the dense clumps catalogued in the simulations, compared with the star-forming dense clumps in \citet{wu:2010.clumps} and \citet{heyer:2016.clumps}. The grey contour encloses the $\pm2\sigma$ contours of clump size at a given mass.}
\label{fig:Clump_M_v_R}
\end{figure}

\begin{figure}
\includegraphics[width=\columnwidth]{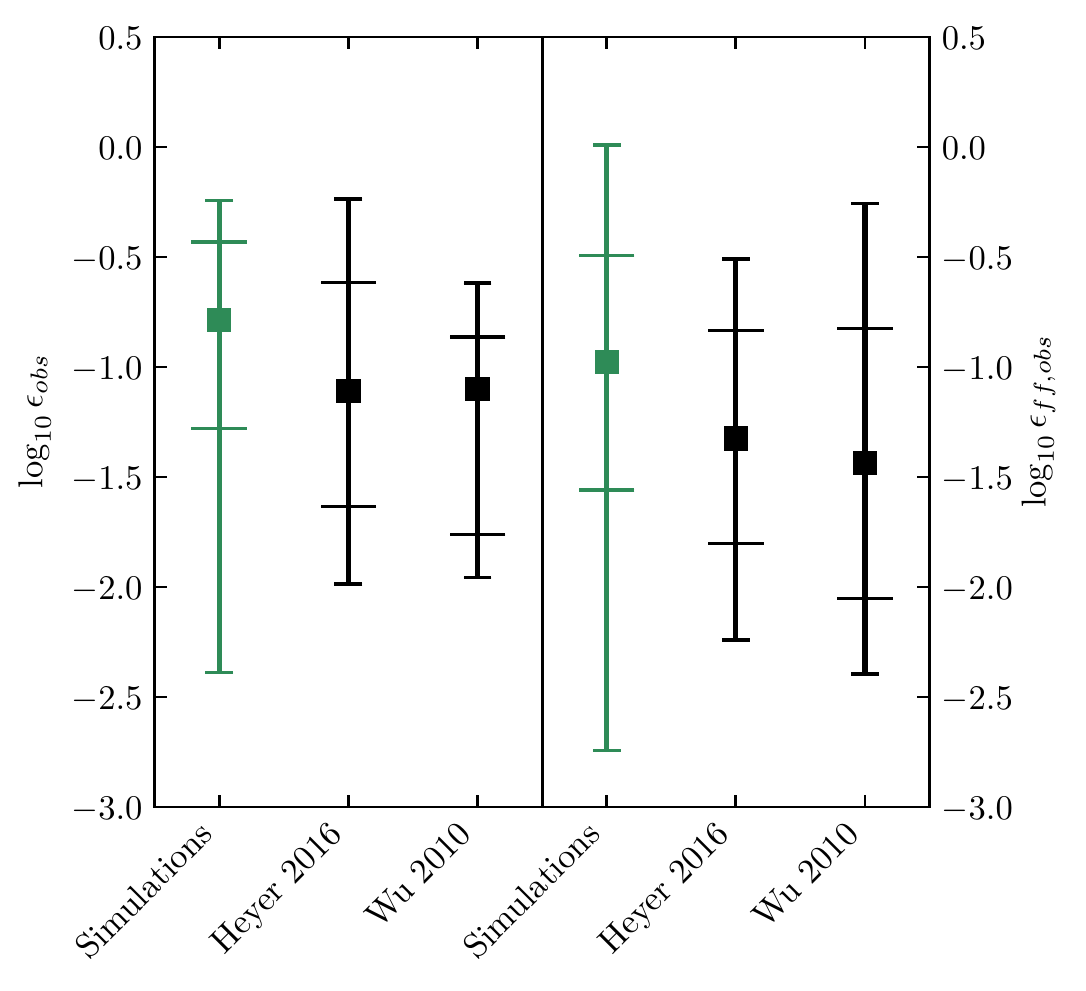}
\caption{Comparison of the distributions (medians, $\pm \sigma$, and $\pm 2\sigma$ quantiles) of $\epsilon_{obs}$ and $\epsilon_{ff,obs}$ for dense ($n_{{\rm H}_2} > 10^4{\rm cm}^{-3}$) clumps in the simulations with the star-forming dense clump datasets from \citet{wu:2010.clumps} and \citet{heyer:2016.clumps}. We plot $\epsilon_{ff,obs}$ as calculated from the same procedure as \citet{heyer:2016.clumps}, which assumes a fixed tracer-identified SF timescale $\tau_{tr}=0.5{\rm Myr}$.} 
\label{fig:denseclumps}
\end{figure}
Because only a small fraction of the initial gas mass of a GMC is converted to stars (e.g. Figure \ref{fig:tff_sfe}), one expects that the densest regions of molecular clouds are the most tightly correlated with star formation activity. Thus far we have examined the behaviour of SFE observables derived from integrated quantities for entire GMCs, and found that molecular cloud evolution under the influence of feedback from massive stars is a satisfactory explanation for the observed ranges of cloud SFEs. 
We now 
examine the properties of dense subregions of molecular clouds and determine whether 
our
model can also explain observations on this smaller scale. We shall consider both observations pertaining to the total gas mass above a certain density threshold within a cloud, and the properties of individual dense clumps.

\subsection{SFR versus gas mass above a 2D/3D density threshold}

In nearby star-forming regions, a proportional relation was found between the SFR and the gas mass at above a certain K-band extinction threshold by \citet{lada:2010.gmcs}:
\begin{equation}
{\rm SFR} = 4.6\times 10^{-8} \left(\frac{M_{0.8}}{M_\odot}\right) \left(M_\odot\,{\rm yr^{-1}}\right),
\label{eq:M08}
\end{equation}
where $M_{0.8}$ is the gas mass of the cloud with $K$-band extinction greater than $0.8{\rm mag}$, corresponding to a gas surface density of $\unit[116]{M_\odot\,pc^{-2}}$. To compare the simulations with this relation, we calculate the K-band extinction of a gas cell by calculating the column density via the ray-tracing method described in \S\ref{section:sfeobs.methods}, but using the same K-band dust opacity assumed in \citet{lada:2010.gmcs} \citep{2009A&A...493..735L}. In the left panels of Figure \ref{fig:M_v_SFR} we plot the average SFR over $\tau_{SF}=2{\rm Myr}$ (similar to the YSO count-inferred SFR in \citealt{lada:2010.gmcs}) as a function of $M_{0.8}$. We find that the clouds simulated with stellar feedback do spend a significant fraction of their lifetime on or near the relation. On the other hand, the simulations without feedback almost always lie $\unit[\sim 1]{dex}$ above the relation. We thus reproduce the finding of \citet{geen:2017} that some form of feedback is necessary to place star-forming clouds on the \citet{lada:2010.gmcs} relation. While it is tempting to then claim that the relation is explained by feedback from massive stars, we caution that this type of mechanism cannot explain the SFR of the lowest-mass star-forming regions considered in \citet{lada:2010.gmcs}, because the stellar masses present are so low that no massive stars are expected to be present. Rather, we have simulated only what would be considered high-mass star forming-systems. Explaining the \citet{lada:2010.gmcs} relation for low-mass systems in terms of feedback may require another mechanism that can moderate star formation in the absence of of massive stars.

\citet{lada:2010.gmcs} further conjectured that $M_{0.8}$ might correspond to the gas mass denser than $n_{\mathrm{H}_2} =\unit[10^4]{cm^{-3}}$, denoted $M_{dense}$. If so, the \citet{lada:2010.gmcs} relation would suggest a simple universal star formation relation:
\begin{equation}
\mathrm{SFR} = \epsilon_{ff} M_{dense}/t_{ff},
\label{eq:Mdense}
\end{equation}
where $t_{ff}=\unit[0.3]{Myr}$ is the freefall time at that density and $\epsilon_{ff} \sim 1\%$. This would roughly agree with the star formation relation suggested by the $L_{IR}-L_{HCN}$ correlation \citep{wu:2005.clumps,wu:2010.clumps, bigiel:2016.empire} under the assumption that HCN emission does actually trace the gas mass of characteristic density $\sim 10^4\mathrm{cm}^{-3}$ (\citealt{krumholz:sf.eff.in.clouds,onus:2018.hcn}, however note recent evidence to the contrary: \citealt{kauffmann:2017.hcn,goldsmith:2017.electron.excitation}). We plot the relation between $M_{dense}$ and the SFR in the rightmost panels of Figure \ref{fig:M_v_SFR}, and find that the simulations lie well above a $\epsilon_{ff} \sim 1\%$ relation, and the relation is steeper than linear. As in \citet{clark.glover:2014} and \citet{geen:2017}, we find a correlation but no general proportionality between $M_{dense}$ and $M_{0.8}$ in the simulations, so Equation \ref{eq:Mdense} does not follow from Equation \ref{eq:M08}. Both with and without feedback, the simulations lie mostly in the range $\epsilon_{ff}=10\%-100\%$, so feedback from massive stars as implemented here does not appear to be sufficient to achieve ``slow" star formation in dense gas.

\subsection{Individual Dense Clumps}
To compare to observations of  individual dense clumps in \citet{wu:2010.clumps} and \citet{heyer:2016.clumps}, we identify contiguous regions with $n_{{\rm H}_2}>10^{4}{\rm cm^{-3}}$ in the simulations, and associate these with sink particles younger than $0.5{\rm Myr}$ found within $2R_{eff}$ of the gas centre of mass, with $R_{eff}$ computed as in \S\ref{section:sfeobs.methods}. We find that the stellar masses associated with the dense clumps are relatively insensitive to the choice of cutoff radius beyond this value because young sink particles are tightly clustered around dense clumps. Within this population we find stellar-mass objects that would be more readily identified as ``cores'' rather than ``clumps''. To make a reasonable comparison with observed dense clumps, we apply a mass cut of $100M_\odot$, which excludes these cores.

The mass-size relation of simulated clumps is compared to observations in Figure \ref{fig:Clump_M_v_R}. We find that in the mass-size plane the clump catalogue from the simulations overlaps most of the dense clumps in \citet{heyer:2016.clumps} and roughly half of those in \citet{wu:2010.clumps}. This supports the interpretation of dense clumps as dense subregions that formed dynamically within a larger molecular gas complex. However, although we have simulated cloud models similar to the most massive Milky Way GMCs, we do not find dense clumps with masses as great as the most massive in either catalogue. This might be due to a genuine missing physical mechanism that might slow down gas consumption in dense clumps, allowing them to live longer and accrete to greater masses. However, resolution effects might also account for the discrepancy: observations with finite spatial and/or spectral resolution would be more likely to lump together 
multiple small clumps into a larger single clump.

In Figure \ref{fig:denseclumps} we plot the distributions for $\epsilon_{obs}$ and $\epsilon_{ff}$ of dense clumps in the simulations compared to observations. We compute $\epsilon_{obs}$ from the total dense gas and stellar mass. We compute $\epsilon_{ff,obs}$ via Eq. \ref{eq:effobs} using the fiducial star formation timescale $\tau_{tr}=0.5{\rm Myr}$ used in \citet{heyer:2016.clumps}. We find a similar amount of scatter to what is observed, which is presumably due to similar effects to what we have found on GMC scales. However, as in the previous subsection, we find efficiencies that are generally greater than what is observed: both $\epsilon_{obs}$ and $\epsilon_{ff,obs}$ for dense clumps are systematically $\unit[\sim 0.3]{dex}$ greater than the observations, which themselves are really upper bounds (see discussion in \ref{sec:compilation}). Moreover, the efficiencies and bulk properties of the observed dense clumps agree well despite the use of different methodologies, and the bulk properties of \citet{heyer:2016.clumps} in particular were derived independently of any assumptions about the characteristic density traced by HCN. It therefore seems quite possible that there is a genuine discrepancy in the efficiency of star formation in dense gas in the simulations: the physics that we have included may not be sufficient to slow down star formation in dense gas down to the levels observed.

\subsection{Possible missing physics}
Assuming that the discrepancy in the SFE of dense gas shown in this section is genuine, and not due to some unknown systematic, there are several pieces of physics neglected here that might affect the clump SFEs:
\begin{itemize}
\item Multiply-scattered IR radiation pressure in the optically-thick limit \citep[e.g][]{krumholz:2012.rad.pressure.rt.instab, davis:2014.rad.pressure.outflows,skinner:2015.ir.molcloud.disrupt,zhang:2017.ir.pressure,tsang:2017.ssc.rp}, which can conceivably become comparable or greater in magnitude to the radiation pressure from direct stellar emission at the gas surface density of these clumps, $\unit[>10^3]{\msun\,pc^{-2}}$.
\item Radiative heating from protostellar accretion, which has been found to be sufficient, and possibly necessary, to set the characteristic mass scale of stars \citep{bate:2009.imf,krumholz:2014.review, guszejnov:2016.imf.feedback,federrath:2017.imf}. The simulations form sink particles of stellar mass, so this may well be dynamically relevant on the scales resolved.
\item Protostellar outflows, which have been shown to be able to slow down star formation in high-resolution periodic box simulations on scales similar to dense clumps \citep{ myers:2014.feedback,federrath:2015.mhd.outflows,cunningham:2018.protostellar.mhd}, 
but have not 
been treated in the present context, in which dense clumps form and disperse dynamically within a larger molecular cloud.
\item Hard-scattering N-body interactions, which would not necessarily bring down the actual SFE, but would reduce the observed SFE if able to dynamically eject YSOs from their natal clumps, as in the classic competitive accretion picture \citep{bonnell:2001.competitive.accretion}. Such interactions depend sensitively upon the relative masses of protostars in a clump, so this effect is only expected to have the correct behaviour in simulations that resolve the IMF self-consistently, which we have not attempted to do here -- a numerically-converged IMF likely requires some subset of the physics mentioned in the above points (see references). In a previous iteration of these simulations, we encountered a bug that caused spurious ejection of sink particles from clumps, and found that the resulting measured SFE was in good agreement with observations. Therefore, the efficient removal of stellar mass in a clump crossing time might reduce the SFE to observed levels.
\end{itemize}

\section{Interpretation of the Properties of Star-Forming GMCs}
\begin{figure}
\includegraphics[width=\columnwidth]{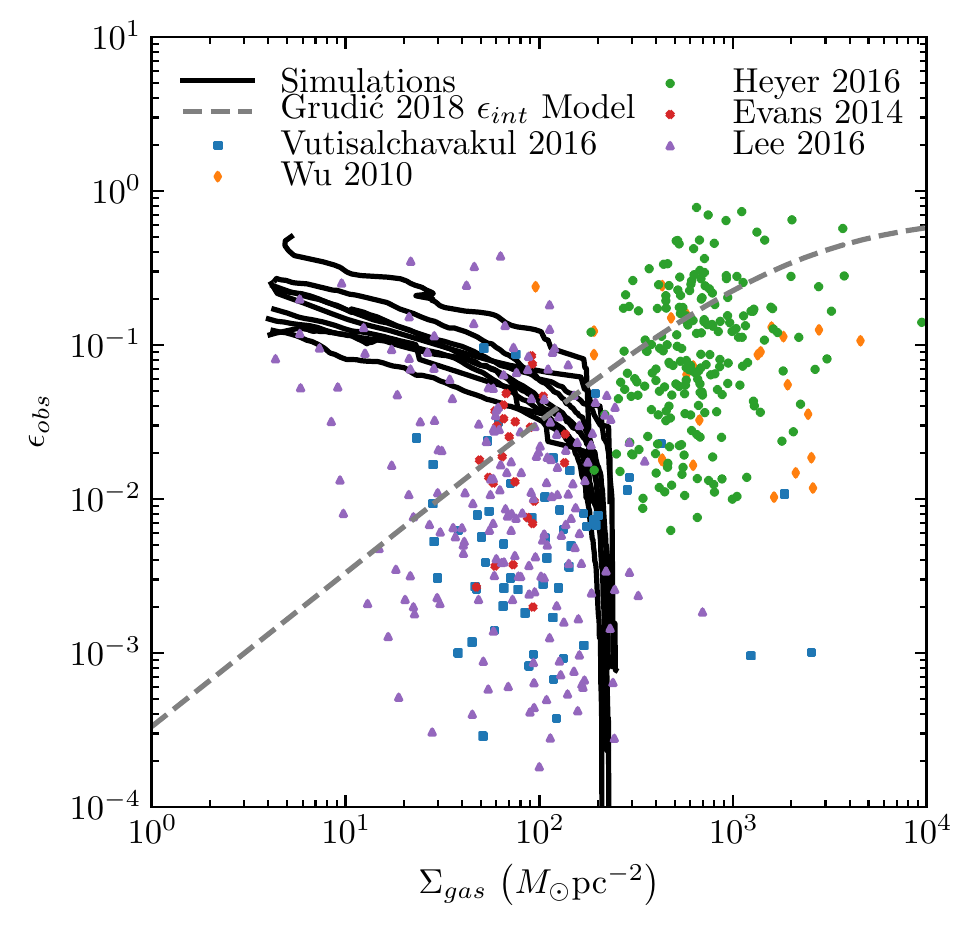}
\caption{Tracks traced out by the simulations in the $\Sigma_{gas}-\epsilon_{obs}$ plane compared with data from star-forming Milky Way clouds and clumps (\S\ref{sec:compilation}) and the \citetalias{grudic:2016.sfe} theory for the dependence of $\epsilon_{int}$ upon $\Sigma_{gas}$. For consistency, $\Sigma_{gas}$ is computed using the effective $R_{eff}$ (see \ref{section:sfeobs.methods}), which is less than the nominal bounding radius $R$ and thus gives a somewhat greater surface density than the nominal $M/\pi R^2$. Both the observed SFEs and surface densities evolve throughout the GMC lifetime, which scatters the data. The prediction of GMC evolution subject to stellar feedback is therefore not a functional dependence of $\epsilon_{int}$ on $\Sigma_{gas}$, but rather a complicated joint probability distribution function that will depend upon the statistics of the underlying cloud population. This complicates the task of discerning true systematic scalings from the observational data.}
\label{fig:tracks}
\end{figure}



\subsection{Does SFE scale with cloud properties?}
An objective of star formation theory is to use observations of star-forming clouds to gain 
insights about the underlying physical mechanisms that determine their evolution and lead to star formation. Constraints can be obtained by comparing observations to the various theories of turbulence-regulated and feedback-regulated star formation, which make specific predictions for $\epsilon_{int}$ and $\epsilon_{ff}$ in terms of the bulk properties of GMCs, such as $M$, $R$, and $\alpha_{vir}$. For example, hydrodynamics simulations with stellar feedback generally predict $\epsilon_{int}$ to scale in some manner with escape velocity, density, or surface density. \citetalias{grudic:2016.sfe} pointed out that the median $\epsilon_{obs}$ appears to scale by a factor of $\sim 10$ over the surface density range separating Milky Way GMCs and dense clumps, which is also roughly a factor of 10 (Figure \ref{fig:tracks}). However, so far such SFE scalings have not yet been conclusively demonstrated in observational studies within a single population of homogeneously-catalogued gas structures. Our simulations suggest that this may be due, at least in part, to the fact that the observable quantities predicted by theory vary in a complex manner that complicates the comparison of theoretical models with observations. 

The observed $M_{mol}$, $R_{eff}$ (and hence $\Sigma_{gas}$) and $\epsilon_{obs}$ of a star-forming GMCs will all vary by orders of magnitude throughout the cloud lifetime (see Figure \ref{fig:tracks}), so numerical simulations of star formation with stellar feedback from massive stars predict a large spread in $\epsilon_{obs}$ in $\epsilon_{ff}$. Therefore, $\epsilon_{obs}$ cannot be treated as a one-to-one function of the bulk cloud properties. To properly investigate possible SFE scalings due to feedback-regulated star formation, it is necessary to model the posterior distribution of predicted ${\it observable}$ GMC properties, which is very broad in both the space of GMC bulk properties and in observed SFE. 
Furthermore, because a GMC will begin its star-forming process below the threshold of detectable massive star formation, and end its evolution in catastrophic dispersal only to be found below the detectable threshold of molecular emission brightness,
some understanding of the selection function is also important to model the likelihood. 
Failing this, any true underlying scalings in SFE with cloud properties can easily be obscured or fit incorrectly. 
Although the simulations equip us with some idea of the likelihood function for a given cloud model, fitting to data from a population of clouds makes it necessary to forward-model the statistics of the cloud parameters within an entire galactic GMC population, which is beyond the scope of this work.

\subsection{Turbulence regulation vs. feedback regulation}
The results of this work do permit certain definite conclusions about the physics governing molecular cloud evolution. We have shown that our simulated GMC models do predict the observed distribution of SFEs of Milky Way GMCs in some detail (Figure \ref{fig:deltaquantiles}), and are able to reproduce the observed anticorrelation of $\epsilon_{obs}$ with total cloud mass \citep{lee:2016.gmc.eff}. \citet{lee:2016.gmc.eff} showed that turbulence-regulated SFE theories derived from the log-normal density PDF and a gravitational collapse criterion \citep[e.g.][]{km2005,hennebelle:2011.time.dept.imf.eps,pn2011} predict neither of these features, generally predicting a positive correlation of SFE with cloud mass with much less scatter than is observed. We therefore favour the model of dynamic star formation in feedback-disrupted GMCs as an explanation for the observed properties of star-forming GMCs in the Milky Way. The observed scatter is due to the large variation in observable gas and stellar mass throughout the cloud lifetime, both due to a dynamic SFR during the initial collapse phase and the depletion of stellar and molecular gas tracers due to stellar evolution and cloud disruption. Furthermore, the normalization of the SFE is not due to regulation by turbulence, but rather stellar feedback from massive stars setting the stellar mass that can be formed before star formation ceases. This is not to minimize the importance of turbulence in the dynamics of star-forming clouds, which is self-evident. Rather, the specific predictions of analytic theories that assume the properties of statistically-stationary, non-self-gravitating turbulence fail to capture the full dynamics of self-gravitating clouds subject to the effects of feedback.


\section{Summary}
In this work, we have presented MHD simulations intended to directly model star-forming GMCs in the Milky Way, accounting for the stellar feedback mechanisms due to massive stars: stellar winds, supernova explosions, and radiation, including the effects of photon momentum in multiple bands, and heating mechanisms due to UV photons. From these simulations we have arrived at several conclusions about the nature of local star-forming molecular clouds:
\begin{itemize}
\item When the effects of magnetic fields and feedback from massive stars are included, the simulations predict an dynamically-rising star formation rate in molecular clouds (Figure \ref{fig:tff_sfe}), as predicted analytically \citep{murray.chang:2015} and found in previous works that considered only gravity and isothermal (magneto-) hydrodynamics \citep{padoan:2012.sfe, federrath:2012.sfr.vs.model.turb.boxes, lee:2015.gravoturbulence,murray:2017} or a different subset of relevant feedback mechanisms \citep{raskutti:2016.gmcs, vazquez:2017,geen:2017}. After this initial growth phase, stellar feedback eventually causes the SFR to level off and drop to 0 as the molecular cloud is disrupted.
\item The simulations predict a normalization and spread in the observed SFEs that is reasonably consistent with those of observed Milky Way GMC SFEs. The diversity in the measured SFE of molecular clouds in the Milky Way is similar to the range of SFE values that is measured across the star-forming lifetime of a single molecular cloud subject to stellar feedback. This stands in contrast to quasi-static models of molecular cloud evolution \citep{zuckerman1974,krumholz.matzner.mckee:2006}, where molecular cloud properties vary on timescales longer than a cloud free-fall time.
\item According to the above interpretation of the SFE spread, very large ($>10\%$) or very small ($<0.1\%$) observed SFEs in individual clouds do not imply that GMCs actually exhibit such great variation in the fraction of their mass that they convert to stars, because these correspond only indirectly to the true SFE values. The true SFE variation could actually be quite small, and the observed scatter would still be observed. Because these effects are a consequence of nonlinear molecular cloud evolution subject to the interplay of feedback, gravity, and hydrodynamics, we concur with \citet{lee:2016.gmc.eff} that theories invoking gravity and turbulence alone cannot explain the observed range of SFE.
\item The observed trend of decreasing $\epsilon_{obs}$ with cloud mass \citep{murray:2010.sfe.mw.gmc,lee:2016.gmc.eff} can be understood as an observational effect arising from the use of only recently-formed stars as a tracer of stellar mass, which underestimates the total stellar mass formed in more massive clouds, which have longer lifetimes. The true SFE in the simulations has no strong trend with mass.
\item We examined the relations between gas mass above 2D and 3D density thresholds in the simulations and the SFR. Simulations lie on the \citet{lada:2010.gmcs} $M_{0.8}-\mathrm{SFR}$ relation (Equation \ref{eq:M08}) if and only if stellar feedback is included, however this does not explain the relation in low-mass star-forming regions where massive stars are absent. However, with or without feedback, the simulations lie well above the proposed corresponding 3D relation for dense ($>10^4\mathrm{cm}^{-3}$) gas (Equation \ref{eq:Mdense}) that assumes that $M_{0.8}=M_{dense}$. This is in agreement with previous simulation work showing that $M_{0.8} \neq M_{dense}$ in general \citep{clark.glover:2014,geen:2017}.
\item We identify contiguous regions of dense ($>10^4\,{\rm cm}^{-3}$) gas within the simulated GMCs with observed dense clumps \citep{wu:2010.clumps,heyer:2016.clumps}, and find that their bulk properties are mostly in good agreement with observed clumps, except for a dearth of predicted clumps more massive than $\unit[3000]{M_\odot}$. We measure the clump SFE in a manner replicating observation techniques and find that $\epsilon_{obs}$ and $\epsilon_{ff,obs}$ are both systematically $\unit[0.3]{dex}$ greater than is observed. It is possible that feedback from main sequence massive stars is insufficient to bring the SFE of dense gas down to the levels observed. 
\end{itemize}

We can identify several avenues for further progress on this problem. Our simulations relied upon ad-hoc initial conditions generated by stirring supersonic MHD turbulence and then ``switching on'' gravity. This is fairly artificial, because in a real galaxy it is likely that gravity plays a role in the actual formation of the molecular cloud and the generation of its turbulent motions. Furthermore, the extent to which even gravitationally-bound clouds can be treated as isolated objects is not well-established, even if they only survive for roughly a free-fall time as they do in the simulations. To address these questions, future studies should account for the greater galactic context of molecular cloud formation and dispersal.

For purposes of determining the stellar feedback budget, we made the single phenomenological assumption of a \citet{kroupa:imf} IMF, which was sampled within sink particles according to the simple prescription of \citet{su:2017.discreteness}. While the universality of the IMF across most Milky Way environments is reasonably well-established \citep{offner:2014.imf.review}, in reality the IMF must somehow emerge from the dynamics of star-forming clouds. Therefore, the manner in which we have decoupled the formation of massive stars from the actual local cloud dynamics is not fully self-consistent. If special conditions are actually necessary for massive star formation, the simulations would not capture the effect and might overestimate feedback. A fully self-consistent molecular cloud simulation with sufficient resolution and physics to {\it predict} the IMF would be necessary to validate our approach. This presents a challenging resolution requirement due to the factor of $\sim 10^6$ disparity in mass scale between the average molecular cloud and the average star. However, recent cloud-collapse simulations without stellar feedback have managed to scale to a relative mass resolution of $<10^{-8}$ \citep{guszejnov:2018.isothermal}, demonstrating that GMC simulations with such a dynamic range are becoming possible.

Lastly, we caution that many conclusions about molecular clouds can be sensitive to the definition of a cloud. For instance, there is no one well-defined way to decompose a CO emission map into clouds, because the ISM exhibits substructure on all scales from the galactic scale height to individual stars, with no obvious preferred intermediate scale. It is likely that GMCs do exist as well-defined dynamical entities in the sense that they may be identified with the largest self-gravitating gas structures within a galaxy \citep{rosolowsky:2008.gmcs,hopkins:2012.excursion.set}, but these do not necessarily correspond to observationally-catalogued GMCs on a one-to-one basis. A more sensitive quantitative comparison of simulated SFEs with observations than we have presented here should account for this by applying the same observational cloud decomposition and cross-correlation algorithms to mock observations.

\section*{Acknowledgements}
We thank Neal J. Evans II, Mark Krumholz, Diederik Kruijssen, Marta Reina-Campos, and Sharon Meidt for enlightening discussions that informed and motivated this work. We also thank Shea Garrison-Kimmel and Alex Gurvich for helpful suggestions for data presentation and visualization. Support for MYG and PFH was provided by a James A Cullen Memorial Fellowship, an Alfred P. Sloan Research Fellowship, NSF Collaborative Research Grant \#1715847 and CAREER grant \#1455342, and NASA grants NNX15AT06G, JPL 1589742, 17-ATP17-0214. CAFG was supported by NSF through grants AST-1412836, AST-1517491, AST-1715216, and CAREER award AST-1652522, by NASA through grant NNX15AB22G, and by a Cottrell Scholar Award from the Research Corporation for Science Advancement. NM acknowledges the support of the Natural Sciences and Engineering Research Council of Canada (NSERC). This research was undertaken, in part, thanks to funding from the Canada Research Chairs program. NM's work was performed in part at the Aspen Center for Physics, which is supported by National Science Foundation grant PHY-1607611. Numerical calculations were run on the Caltech compute cluster ``Wheeler,'' allocations from XSEDE TG-AST130039 and PRAC NSF.1713353 (awards OCI-0725070 and ACI-1238993) supported by the NSF, and NASA HEC SMD-16-7592. 
This research has made use of use of NASA's Astrophysics Data System, {\texttt ipython} \citep{ipython}, {\texttt numpy}, {\texttt scipy} \citep{scipy}, and {\texttt matplotlib} \citep{matplotlib}.




\bibliographystyle{mnras}
\bibliography{master} 

\begin{thebibliography}{}
\makeatletter
\relax
\def\mn@urlcharsother{\let\do\@makeother \do\$\do\&\do\#\do\^\do\_\do\%\do\~}
\def\mn@doi{\begingroup\mn@urlcharsother \@ifnextchar [ {\mn@doi@}
  {\mn@doi@[]}}
\def\mn@doi@[#1]#2{\def\@tempa{#1}\ifx\@tempa\@empty \href
  {http://dx.doi.org/#2} {doi:#2}\else \href {http://dx.doi.org/#2} {#1}\fi
  \endgroup}
\def\mn@eprint#1#2{\mn@eprint@#1:#2::\@nil}
\def\mn@eprint@arXiv#1{\href {http://arxiv.org/abs/#1} {{\tt arXiv:#1}}}
\def\mn@eprint@dblp#1{\href {http://dblp.uni-trier.de/rec/bibtex/#1.xml}
  {dblp:#1}}
\def\mn@eprint@#1:#2:#3:#4\@nil{\def\@tempa {#1}\def\@tempb {#2}\def\@tempc
  {#3}\ifx \@tempc \@empty \let \@tempc \@tempb \let \@tempb \@tempa \fi \ifx
  \@tempb \@empty \def\@tempb {arXiv}\fi \@ifundefined
  {mn@eprint@\@tempb}{\@tempb:\@tempc}{\expandafter \expandafter \csname
  mn@eprint@\@tempb\endcsname \expandafter{\@tempc}}}

\bibitem[\protect\citeauthoryear{{Ballesteros-Paredes}, {Hartmann},
  {V{\'a}zquez-Semadeni}, {Heitsch}  \&
  {Zamora-Avil{\'e}s}}{{Ballesteros-Paredes}
  et~al.}{2011}]{ballesteros.paredes:2011}
{Ballesteros-Paredes} J.,  {Hartmann} L.~W.,  {V{\'a}zquez-Semadeni} E.,
  {Heitsch} F.,   {Zamora-Avil{\'e}s} M.~A.,  2011, \mn@doi [\mnras]
  {10.1111/j.1365-2966.2010.17657.x}, \href
  {http://adsabs.harvard.edu/abs/2011MNRAS.411...65B} {411, 65}

\bibitem[\protect\citeauthoryear{{Bate}}{{Bate}}{2009}]{bate:2009.imf}
{Bate} M.~R.,  2009, \mn@doi [\mnras] {10.1111/j.1365-2966.2008.14165.x}, \href
  {http://adsabs.harvard.edu/abs/2009MNRAS.392.1363B} {392, 1363}

\bibitem[\protect\citeauthoryear{{Bauer} \& {Springel}}{{Bauer} \&
  {Springel}}{2012}]{bauer.springel:2012}
{Bauer} A.,  {Springel} V.,  2012, \mn@doi [\mnras]
  {10.1111/j.1365-2966.2012.21058.x}, \href
  {http://adsabs.harvard.edu/abs/2012MNRAS.423.2558B} {423, 2558}

\bibitem[\protect\citeauthoryear{{Baumgardt} \& {Kroupa}}{{Baumgardt} \&
  {Kroupa}}{2007}]{baumgardt.kroupa:2007}
{Baumgardt} H.,  {Kroupa} P.,  2007, \mn@doi [\mnras]
  {10.1111/j.1365-2966.2007.12209.x}, \href
  {http://adsabs.harvard.edu/abs/2007MNRAS.380.1589B} {380, 1589}

\bibitem[\protect\citeauthoryear{{Bertoldi} \& {McKee}}{{Bertoldi} \&
  {McKee}}{1992}]{bertoldi.mckee}
{Bertoldi} F.,  {McKee} C.~F.,  1992, \mn@doi [\apj] {10.1086/171638}, \href
  {http://adsabs.harvard.edu/abs/1992ApJ...395..140B} {395, 140}

\bibitem[\protect\citeauthoryear{{Bigiel} et~al.,}{{Bigiel}
  et~al.}{2016}]{bigiel:2016.empire}
{Bigiel} F.,  et~al., 2016, \mn@doi [\apjl] {10.3847/2041-8205/822/2/L26},
  \href {http://adsabs.harvard.edu/abs/2016ApJ...822L..26B} {822, L26}

\bibitem[\protect\citeauthoryear{{Bolatto}, {Leroy}, {Rosolowsky}, {Walter}  \&
  {Blitz}}{{Bolatto} et~al.}{2008}]{bolatto:2008.gmc.properties}
{Bolatto} A.~D.,  {Leroy} A.~K.,  {Rosolowsky} E.,  {Walter} F.,   {Blitz} L.,
  2008, \mn@doi [\apj] {10.1086/591513}, \href
  {http://adsabs.harvard.edu/abs/2008ApJ...686..948B} {686, 948}

\bibitem[\protect\citeauthoryear{{Bolatto}, {Wolfire}  \& {Leroy}}{{Bolatto}
  et~al.}{2013}]{bolatto:2013.xco}
{Bolatto} A.~D.,  {Wolfire} M.,   {Leroy} A.~K.,  2013, \mn@doi [\araa]
  {10.1146/annurev-astro-082812-140944}, \href
  {http://adsabs.harvard.edu/abs/2013ARA%26A..51..207B} {51, 207}

\bibitem[\protect\citeauthoryear{{Bonnell}, {Bate}, {Clarke}  \&
  {Pringle}}{{Bonnell} et~al.}{2001}]{bonnell:2001.competitive.accretion}
{Bonnell} I.~A.,  {Bate} M.~R.,  {Clarke} C.~J.,   {Pringle} J.~E.,  2001,
  \mn@doi [\mnras] {10.1046/j.1365-8711.2001.04270.x}, \href
  {http://adsabs.harvard.edu/abs/2001MNRAS.323..785B} {323, 785}

\bibitem[\protect\citeauthoryear{{Braine}, {Rosolowsky}, {Gratier}, {Corbelli}
  \& {Schuster}}{{Braine} et~al.}{2018}]{braine:2018.gmcs}
{Braine} J.,  {Rosolowsky} E.,  {Gratier} P.,  {Corbelli} E.,   {Schuster} K.,
  2018, preprint, \href {http://adsabs.harvard.edu/abs/2018arXiv180104171B} {}
  (\mn@eprint {arXiv} {1801.04171})

\bibitem[\protect\citeauthoryear{{Burkhart}}{{Burkhart}}{2018}]{burkhart:2018.density.pdf}
{Burkhart} B.,  2018, preprint, \href
  {http://adsabs.harvard.edu/abs/2018arXiv180105428B} {} (\mn@eprint {arXiv}
  {1801.05428})

\bibitem[\protect\citeauthoryear{{Clark} \& {Glover}}{{Clark} \&
  {Glover}}{2014}]{clark.glover:2014}
{Clark} P.~C.,  {Glover} S.~C.~O.,  2014, \mn@doi [\mnras]
  {10.1093/mnras/stu1589}, \href
  {http://adsabs.harvard.edu/abs/2014MNRAS.444.2396C} {444, 2396}

\bibitem[\protect\citeauthoryear{{Col{\'{\i}}n}, {V{\'a}zquez-Semadeni}  \&
  {G{\'o}mez}}{{Col{\'{\i}}n} et~al.}{2013}]{colin:2013}
{Col{\'{\i}}n} P.,  {V{\'a}zquez-Semadeni} E.,   {G{\'o}mez} G.~C.,  2013,
  \mn@doi [\mnras] {10.1093/mnras/stt1409}, \href
  {http://adsabs.harvard.edu/abs/2013MNRAS.435.1701C} {435, 1701}

\bibitem[\protect\citeauthoryear{{Cunningham}, {Krumholz}, {McKee}  \&
  {Klein}}{{Cunningham} et~al.}{2018}]{cunningham:2018.protostellar.mhd}
{Cunningham} A.~J.,  {Krumholz} M.~R.,  {McKee} C.~F.,   {Klein} R.~I.,  2018,
  \mn@doi [\mnras] {10.1093/mnras/sty154}, \href
  {http://adsabs.harvard.edu/abs/2018MNRAS.476..771C} {476, 771}

\bibitem[\protect\citeauthoryear{{Dale}}{{Dale}}{2015}]{dale:2015.review}
{Dale} J.~E.,  2015, \mn@doi [\nar] {10.1016/j.newar.2015.06.001}, \href
  {http://adsabs.harvard.edu/abs/2015NewAR..68....1D} {68, 1}

\bibitem[\protect\citeauthoryear{{Dale}}{{Dale}}{2017}]{dale:2017}
{Dale} J.~E.,  2017, \mn@doi [\mnras] {10.1093/mnras/stx028}, \href
  {http://adsabs.harvard.edu/abs/2017MNRAS.467.1067D} {467, 1067}

\bibitem[\protect\citeauthoryear{{Dale}, {Ercolano}  \& {Bonnell}}{{Dale}
  et~al.}{2012}]{dale:2012}
{Dale} J.~E.,  {Ercolano} B.,   {Bonnell} I.~A.,  2012, \mn@doi [\mnras]
  {10.1111/j.1365-2966.2012.21205.x}, \href
  {http://adsabs.harvard.edu/abs/2012MNRAS.424..377D} {424, 377}

\bibitem[\protect\citeauthoryear{{Dale}, {Ngoumou}, {Ercolano}  \&
  {Bonnell}}{{Dale} et~al.}{2013}]{dale:2013.momwinds}
{Dale} J.~E.,  {Ngoumou} J.,  {Ercolano} B.,   {Bonnell} I.~A.,  2013, \mn@doi
  [\mnras] {10.1093/mnras/stt1822}, \href
  {http://adsabs.harvard.edu/abs/2013MNRAS.436.3430D} {436, 3430}

\bibitem[\protect\citeauthoryear{{Dale}, {Ngoumou}, {Ercolano}  \&
  {Bonnell}}{{Dale} et~al.}{2014}]{dale:2014}
{Dale} J.~E.,  {Ngoumou} J.,  {Ercolano} B.,   {Bonnell} I.~A.,  2014, \mn@doi
  [\mnras] {10.1093/mnras/stu816}, \href
  {http://adsabs.harvard.edu/abs/2014MNRAS.442..694D} {442, 694}

\bibitem[\protect\citeauthoryear{{Davis}, {Jiang}, {Stone}  \&
  {Murray}}{{Davis} et~al.}{2014}]{davis:2014.rad.pressure.outflows}
{Davis} S.~W.,  {Jiang} Y.-F.,  {Stone} J.~M.,   {Murray} N.,  2014, \apj, in
  press, arXiv:1403.1874, \href
  {http://adsabs.harvard.edu/abs/2014arXiv1403.1874D} {}

\bibitem[\protect\citeauthoryear{{Dobbs} \& {Pringle}}{{Dobbs} \&
  {Pringle}}{2013}]{dobbs:2013}
{Dobbs} C.~L.,  {Pringle} J.~E.,  2013, \mn@doi [\mnras]
  {10.1093/mnras/stt508}, \href
  {http://adsabs.harvard.edu/abs/2013MNRAS.432..653D} {432, 653}

\bibitem[\protect\citeauthoryear{{Elmegreen} \& {Clemens}}{{Elmegreen} \&
  {Clemens}}{1985}]{1985ApJ...294..523E}
{Elmegreen} B.~G.,  {Clemens} C.,  1985, \mn@doi [\apj] {10.1086/163320}, \href
  {http://adsabs.harvard.edu/abs/1985ApJ...294..523E} {294, 523}

\bibitem[\protect\citeauthoryear{{Evans} II et~al.,}{{Evans}
  et~al.}{2009}]{evans:2009.sfe}
{Evans} II N.~J.,  et~al., 2009, \mn@doi [\apjs] {10.1088/0067-0049/181/2/321},
  \href {http://adsabs.harvard.edu/abs/2009ApJS..181..321E} {181, 321}

\bibitem[\protect\citeauthoryear{{Evans}, {Heiderman}  \&
  {Vutisalchavakul}}{{Evans} et~al.}{2014}]{evans:2014.sfe}
{Evans} II N.~J.,  {Heiderman} A.,   {Vutisalchavakul} N.,  2014, \mn@doi
  [\apj] {10.1088/0004-637X/782/2/114}, \href
  {http://adsabs.harvard.edu/abs/2014ApJ...782..114E} {782, 114}

\bibitem[\protect\citeauthoryear{{Fall}, {Krumholz}  \& {Matzner}}{{Fall}
  et~al.}{2010}]{fall:2010.sf.eff.vs.surfacedensity}
{Fall} S.~M.,  {Krumholz} M.~R.,   {Matzner} C.~D.,  2010, \mn@doi [\apjl]
  {10.1088/2041-8205/710/2/L142}, \href
  {http://adsabs.harvard.edu/abs/2010ApJ...710L.142F} {710, L142}

\bibitem[\protect\citeauthoryear{{Faucher-Gigu{\`e}re}, {Quataert}  \&
  {Hopkins}}{{Faucher-Gigu{\`e}re} et~al.}{2013}]{cafg:sf.fb.reg.kslaw}
{Faucher-Gigu{\`e}re} C.-A.,  {Quataert} E.,   {Hopkins} P.~F.,  2013, \mn@doi
  [\mnras] {10.1093/mnras/stt866}, \href
  {http://adsabs.harvard.edu/abs/2013MNRAS.433.1970F} {433, 1970}

\bibitem[\protect\citeauthoryear{{Federrath}}{{Federrath}}{2015}]{federrath:2015.mhd.outflows}
{Federrath} C.,  2015, \mn@doi [\mnras] {10.1093/mnras/stv941}, \href
  {http://adsabs.harvard.edu/abs/2015MNRAS.450.4035F} {450, 4035}

\bibitem[\protect\citeauthoryear{{Federrath} \& {Klessen}}{{Federrath} \&
  {Klessen}}{2012}]{federrath:2012.sfr.vs.model.turb.boxes}
{Federrath} C.,  {Klessen} R.~S.,  2012, \mn@doi [\apj]
  {10.1088/0004-637X/761/2/156}, \href
  {http://adsabs.harvard.edu/abs/2012arXiv1209.2856F} {761, 156}

\bibitem[\protect\citeauthoryear{{Federrath}, {Banerjee}, {Clark}  \&
  {Klessen}}{{Federrath} et~al.}{2010}]{Federrath_2010_sink_particle}
{Federrath} C.,  {Banerjee} R.,  {Clark} P.~C.,   {Klessen} R.~S.,  2010,
  \mn@doi [\apj] {10.1088/0004-637X/713/1/269}, \href
  {http://adsabs.harvard.edu/abs/2010ApJ...713..269F} {713, 269}

\bibitem[\protect\citeauthoryear{{Federrath}, {Schober}, {Bovino}  \&
  {Schleicher}}{{Federrath} et~al.}{2014}]{federrath:supersonic.turb.dynamo}
{Federrath} C.,  {Schober} J.,  {Bovino} S.,   {Schleicher} D.~R.~G.,  2014,
  \mn@doi [\apjl] {10.1088/2041-8205/797/2/L19}, \href
  {http://adsabs.harvard.edu/abs/2014ApJ...797L..19F} {797, L19}

\bibitem[\protect\citeauthoryear{{Federrath}, {Krumholz}  \&
  {Hopkins}}{{Federrath} et~al.}{2017}]{federrath:2017.imf}
{Federrath} C.,  {Krumholz} M.,   {Hopkins} P.~F.,  2017, in Journal of Physics
  Conference Series. p. 012007, \mn@doi{10.1088/1742-6596/837/1/012007}

\bibitem[\protect\citeauthoryear{{Feldmann} \& {Gnedin}}{{Feldmann} \&
  {Gnedin}}{2011}]{feldmann:2011}
{Feldmann} R.,  {Gnedin} N.~Y.,  2011, \mn@doi [\apjl]
  {10.1088/2041-8205/727/1/L12}, \href
  {http://adsabs.harvard.edu/abs/2011ApJ...727L..12F} {727, L12}

\bibitem[\protect\citeauthoryear{{Fukui} \& {Kawamura}}{{Fukui} \&
  {Kawamura}}{2010}]{2010ARA&A..48..547F}
{Fukui} Y.,  {Kawamura} A.,  2010, \mn@doi [\araa]
  {10.1146/annurev-astro-081309-130854}, \href
  {http://adsabs.harvard.edu/abs/2010ARA%26A..48..547F} {48, 547}

\bibitem[\protect\citeauthoryear{{Gammie} \& {Ostriker}}{{Gammie} \&
  {Ostriker}}{1996}]{gammie.ostriker:1996.mhd.dissipation}
{Gammie} C.~F.,  {Ostriker} E.~C.,  1996, \mn@doi [\apj] {10.1086/177556},
  \href {http://adsabs.harvard.edu/abs/1996ApJ...466..814G} {466, 814}

\bibitem[\protect\citeauthoryear{{Gao} \& {Solomon}}{{Gao} \&
  {Solomon}}{2004}]{gao.solomon:2004a.hcn}
{Gao} Y.,  {Solomon} P.~M.,  2004, \mn@doi [\apjs] {10.1086/383003}, \href
  {http://adsabs.harvard.edu/abs/2004ApJS..152...63G} {152, 63}

\bibitem[\protect\citeauthoryear{{Gavagnin}, {Bleuler}, {Rosdahl}  \&
  {Teyssier}}{{Gavagnin} et~al.}{2017}]{gavagnin:2017.rhd.cluster.formation}
{Gavagnin} E.,  {Bleuler} A.,  {Rosdahl} J.,   {Teyssier} R.,  2017, \mn@doi
  [\mnras] {10.1093/mnras/stx2222}, \href
  {http://adsabs.harvard.edu/abs/2017MNRAS.472.4155G} {472, 4155}

\bibitem[\protect\citeauthoryear{{Geen}, {Soler}  \& {Hennebelle}}{{Geen}
  et~al.}{2017}]{geen:2017}
{Geen} S.,  {Soler} J.~D.,   {Hennebelle} P.,  2017, \mn@doi [\mnras]
  {10.1093/mnras/stx1765}, \href
  {http://adsabs.harvard.edu/abs/2017MNRAS.471.4844G} {471, 4844}

\bibitem[\protect\citeauthoryear{{Goldsmith} \& {Kauffmann}}{{Goldsmith} \&
  {Kauffmann}}{2017}]{goldsmith:2017.electron.excitation}
{Goldsmith} P.~F.,  {Kauffmann} J.,  2017, \mn@doi [\apj]
  {10.3847/1538-4357/aa6f12}, \href
  {http://adsabs.harvard.edu/abs/2017ApJ...841...25G} {841, 25}

\bibitem[\protect\citeauthoryear{{Grudi{\'c}}, {Hopkins},
  {Faucher-Gigu{\`e}re}, {Quataert}, {Murray}  \& {Kere{\v s}}}{{Grudi{\'c}}
  et~al.}{2018}]{grudic:2016.sfe}
{Grudi{\'c}} M.~Y.,  {Hopkins} P.~F.,  {Faucher-Gigu{\`e}re} C.-A.,  {Quataert}
  E.,  {Murray} N.,   {Kere{\v s}} D.,  2018, \mn@doi [\mnras]
  {10.1093/mnras/sty035}, \href
  {http://adsabs.harvard.edu/abs/2018MNRAS.475.3511G} {475, 3511}

\bibitem[\protect\citeauthoryear{{Guszejnov}, {Krumholz}  \&
  {Hopkins}}{{Guszejnov} et~al.}{2016}]{guszejnov:2016.imf.feedback}
{Guszejnov} D.,  {Krumholz} M.~R.,   {Hopkins} P.~F.,  2016, \mn@doi [\mnras]
  {10.1093/mnras/stw315}, \href
  {http://adsabs.harvard.edu/abs/2016MNRAS.458..673G} {458, 673}

\bibitem[\protect\citeauthoryear{{Guszejnov}, {Hopkins}, {Grudi{\'c}},
  {Krumholz}  \& {Federrath}}{{Guszejnov}
  et~al.}{2018}]{guszejnov:2018.isothermal}
{Guszejnov} D.,  {Hopkins} P.~F.,  {Grudi{\'c}} M.~Y.,  {Krumholz} M.~R.,
  {Federrath} C.,  2018, \mn@doi [\mnras] {10.1093/mnras/sty1847}, \href
  {http://adsabs.harvard.edu/abs/2018MNRAS.480..182G} {480, 182}

\bibitem[\protect\citeauthoryear{{Heiderman}, {Evans}, {Allen}, {Huard}  \&
  {Heyer}}{{Heiderman} et~al.}{2010}]{heiderman:2010.gmcs}
{Heiderman} A.,  {Evans} II N.~J.,  {Allen} L.~E.,  {Huard} T.,   {Heyer} M.,
  2010, \mn@doi [\apj] {10.1088/0004-637X/723/2/1019}, \href
  {http://adsabs.harvard.edu/abs/2010ApJ...723.1019H} {723, 1019}

\bibitem[\protect\citeauthoryear{{Hennebelle} \& {Chabrier}}{{Hennebelle} \&
  {Chabrier}}{2011a}]{hc2011}
{Hennebelle} P.,  {Chabrier} G.,  2011a, \mn@doi [\apjl]
  {10.1088/2041-8205/743/2/L29}, \href
  {http://adsabs.harvard.edu/abs/2011ApJ...743L..29H} {743, L29}

\bibitem[\protect\citeauthoryear{{Hennebelle} \& {Chabrier}}{{Hennebelle} \&
  {Chabrier}}{2011b}]{hennebelle:2011.time.dept.imf.eps}
{Hennebelle} P.,  {Chabrier} G.,  2011b, \mn@doi [\apjl]
  {10.1088/2041-8205/743/2/L29}, \href
  {http://adsabs.harvard.edu/abs/2011ApJ...743L..29H} {743, L29}

\bibitem[\protect\citeauthoryear{{Heyer}, {Gutermuth}, {Urquhart}, {Csengeri},
  {Wienen}, {Leurini}, {Menten}  \& {Wyrowski}}{{Heyer}
  et~al.}{2016}]{heyer:2016.clumps}
{Heyer} M.,  {Gutermuth} R.,  {Urquhart} J.~S.,  {Csengeri} T.,  {Wienen} M.,
  {Leurini} S.,  {Menten} K.,   {Wyrowski} F.,  2016, \mn@doi [\aap]
  {10.1051/0004-6361/201527681}, \href
  {http://adsabs.harvard.edu/abs/2016A%26A...588A..29H} {588, A29}

\bibitem[\protect\citeauthoryear{{Hills}}{{Hills}}{1980}]{hills:1980}
{Hills} J.~G.,  1980, \mn@doi [\apj] {10.1086/157703}, \href
  {http://adsabs.harvard.edu/abs/1980ApJ...235..986H} {235, 986}

\bibitem[\protect\citeauthoryear{{Hopkins}}{{Hopkins}}{2012}]{hopkins:2012.excursion.set}
{Hopkins} P.~F.,  2012, \mn@doi [\mnras] {10.1111/j.1365-2966.2012.20730.x},
  \href {http://adsabs.harvard.edu/abs/2012MNRAS.423.2016H} {423, 2016}

\bibitem[\protect\citeauthoryear{{Hopkins}}{{Hopkins}}{2015}]{hopkins:gizmo}
{Hopkins} P.~F.,  2015, \mn@doi [\mnras] {10.1093/mnras/stv195}, \href
  {http://adsabs.harvard.edu/abs/2015MNRAS.450...53H} {450, 53}

\bibitem[\protect\citeauthoryear{{Hopkins} \& {Raives}}{{Hopkins} \&
  {Raives}}{2016}]{hopkins:gizmo.mhd}
{Hopkins} P.~F.,  {Raives} M.~J.,  2016, \mn@doi [\mnras]
  {10.1093/mnras/stv2180}, \href
  {http://adsabs.harvard.edu/abs/2016MNRAS.455...51H} {455, 51}

\bibitem[\protect\citeauthoryear{{Hopkins}, {Quataert}  \& {Murray}}{{Hopkins}
  et~al.}{2012}]{hopkins:fb.ism.prop}
{Hopkins} P.~F.,  {Quataert} E.,   {Murray} N.,  2012, \mn@doi [\mnras]
  {10.1111/j.1365-2966.2012.20578.x}, \href
  {http://adsabs.harvard.edu/abs/2012MNRAS.421.3488H} {421, 3488}

\bibitem[\protect\citeauthoryear{{Hopkins}, {Keres}, {Onorbe},
  {Faucher-Giguere}, {Quataert}, {Murray}  \& {Bullock}}{{Hopkins}
  et~al.}{2014}]{hopkins:2013.fire}
{Hopkins} P.~F.,  {Keres} D.,  {Onorbe} J.,  {Faucher-Giguere} C.-A.,
  {Quataert} E.,  {Murray} N.,   {Bullock} J.~S.,  2014, \mn@doi [\mnras]
  {10.1093/mnras/stu1738}, \href
  {http://adsabs.harvard.edu/abs/2013arXiv1311.2073H} {445, 581}

\bibitem[\protect\citeauthoryear{{Hopkins} et~al.,}{{Hopkins}
  et~al.}{2018}]{fire2}
{Hopkins} P.~F.,  et~al., 2018, \mn@doi [\mnras] {10.1093/mnras/sty1690}, \href
  {http://adsabs.harvard.edu/abs/2018MNRAS.480..800H} {480, 800}

\bibitem[\protect\citeauthoryear{{Howard}, {Pudritz}  \& {Harris}}{{Howard}
  et~al.}{2016}]{howard:2016}
{Howard} C.~S.,  {Pudritz} R.~E.,   {Harris} W.~E.,  2016, \mn@doi [\mnras]
  {10.1093/mnras/stw1476}, \href
  {http://adsabs.harvard.edu/abs/2016MNRAS.461.2953H} {461, 2953}

\bibitem[\protect\citeauthoryear{{Howard}, {Pudritz}  \& {Harris}}{{Howard}
  et~al.}{2017}]{howard:2017}
{Howard} C.~S.,  {Pudritz} R.~E.,   {Harris} W.~E.,  2017, \mn@doi [\mnras]
  {10.1093/mnras/stx1363}, \href
  {http://adsabs.harvard.edu/abs/2017MNRAS.470.3346H} {470, 3346}

\bibitem[\protect\citeauthoryear{Hunter}{Hunter}{2007}]{matplotlib}
Hunter J.~D.,  2007, \mn@doi [Computing In Science \& Engineering]
  {10.1109/MCSE.2007.55}, 9, 90

\bibitem[\protect\citeauthoryear{Jones, Oliphant, Peterson  et~al.}{Jones
  et~al.}{2001}]{scipy}
Jones E.,  Oliphant T.,  Peterson P.,   et~al., 2001, {SciPy}: Open source
  scientific tools for {Python}, \url {http://www.scipy.org/}

\bibitem[\protect\citeauthoryear{{Kainulainen}, {Beuther}, {Henning}  \&
  {Plume}}{{Kainulainen} et~al.}{2009}]{kainulainen:2009.density.pdf}
{Kainulainen} J.,  {Beuther} H.,  {Henning} T.,   {Plume} R.,  2009, \mn@doi
  [\aap] {10.1051/0004-6361/200913605}, \href
  {http://adsabs.harvard.edu/abs/2009A%26A...508L..35K} {508, L35}

\bibitem[\protect\citeauthoryear{{Kauffmann}, {Goldsmith}, {Melnick}, {Tolls},
  {Guzman}  \& {Menten}}{{Kauffmann} et~al.}{2017}]{kauffmann:2017.hcn}
{Kauffmann} J.,  {Goldsmith} P.~F.,  {Melnick} G.,  {Tolls} V.,  {Guzman} A.,
  {Menten} K.~M.,  2017, \mn@doi [\aap] {10.1051/0004-6361/201731123}, \href
  {http://adsabs.harvard.edu/abs/2017A%26A...605L...5K} {605, L5}

\bibitem[\protect\citeauthoryear{{Kawamura} et~al.,}{{Kawamura}
  et~al.}{2009}]{2009ApJS..184....1K}
{Kawamura} A.,  et~al., 2009, \mn@doi [\apjs] {10.1088/0067-0049/184/1/1},
  \href {http://adsabs.harvard.edu/abs/2009ApJS..184....1K} {184, 1}

\bibitem[\protect\citeauthoryear{{Kennicutt}}{{Kennicutt}}{1998}]{kennicutt98}
{Kennicutt} Jr. R.~C.,  1998, \mn@doi [\apj] {10.1086/305588}, \href
  {http://adsabs.harvard.edu/cgi-bin/nph-bib_query?bibcode=1998ApJ...498..541K&db_key=AST}
  {498, 541}

\bibitem[\protect\citeauthoryear{{Kim}, {Kim}, {Ostriker}  \& {Skinner}}{{Kim}
  et~al.}{2017}]{kim:2017.rhd}
{Kim} J.-G.,  {Kim} W.-T.,  {Ostriker} E.~C.,   {Skinner} M.~A.,  2017, \mn@doi
  [\apj] {10.3847/1538-4357/aa9b80}, \href
  {http://adsabs.harvard.edu/abs/2017ApJ...851...93K} {851, 93}

\bibitem[\protect\citeauthoryear{{Kim}, {Kim}  \& {Ostriker}}{{Kim}
  et~al.}{2018}]{kim:2018}
{Kim} J.-G.,  {Kim} W.-T.,   {Ostriker} E.~C.,  2018, \mn@doi [\apj]
  {10.3847/1538-4357/aabe27}, \href
  {http://adsabs.harvard.edu/abs/2018ApJ...859...68K} {859, 68}

\bibitem[\protect\citeauthoryear{{Kritsuk}, {Norman}  \& {Wagner}}{{Kritsuk}
  et~al.}{2011}]{kritsuk:2011.density.pdf.power.law}
{Kritsuk} A.~G.,  {Norman} M.~L.,   {Wagner} R.,  2011, \mn@doi [\apjl]
  {10.1088/2041-8205/727/1/L20}, \href
  {http://adsabs.harvard.edu/abs/2011ApJ...727L..20K} {727, L20}

\bibitem[\protect\citeauthoryear{{Kroupa}}{{Kroupa}}{2002}]{kroupa:imf}
{Kroupa} P.,  2002, \mn@doi [Science] {10.1126/science.1067524}, \href
  {http://adsabs.harvard.edu/abs/2002Sci...295...82K} {295, 82}

\bibitem[\protect\citeauthoryear{{Krumholz}}{{Krumholz}}{2014}]{krumholz:2014.review}
{Krumholz} M.~R.,  2014, \mn@doi [\physrep] {10.1016/j.physrep.2014.02.001},
  \href {http://adsabs.harvard.edu/abs/2014PhR...539...49K} {539, 49}

\bibitem[\protect\citeauthoryear{{Krumholz} \& {Gnedin}}{{Krumholz} \&
  {Gnedin}}{2011}]{krumholz:2011.molecular.prescription}
{Krumholz} M.~R.,  {Gnedin} N.~Y.,  2011, \mn@doi [\apj]
  {10.1088/0004-637X/729/1/36}, \href
  {http://adsabs.harvard.edu/abs/2011ApJ...729...36K} {729, 36}

\bibitem[\protect\citeauthoryear{{Krumholz} \& {McKee}}{{Krumholz} \&
  {McKee}}{2005}]{km2005}
{Krumholz} M.~R.,  {McKee} C.~F.,  2005, \mn@doi [\apj] {10.1086/431734}, \href
  {http://adsabs.harvard.edu/abs/2005ApJ...630..250K} {630, 250}

\bibitem[\protect\citeauthoryear{{Krumholz} \& {Tan}}{{Krumholz} \&
  {Tan}}{2007}]{krumholz:sf.eff.in.clouds}
{Krumholz} M.~R.,  {Tan} J.~C.,  2007, \mn@doi [\apj] {10.1086/509101}, \href
  {http://adsabs.harvard.edu/abs/2007ApJ...654..304K} {654, 304}

\bibitem[\protect\citeauthoryear{{Krumholz} \& {Thompson}}{{Krumholz} \&
  {Thompson}}{2012}]{krumholz:2012.rad.pressure.rt.instab}
{Krumholz} M.~R.,  {Thompson} T.~A.,  2012, \mn@doi [\apj]
  {10.1088/0004-637X/760/2/155}, \href
  {http://adsabs.harvard.edu/abs/2012arXiv1203.2926K} {760, 155}

\bibitem[\protect\citeauthoryear{{Krumholz}, {Matzner}  \& {McKee}}{{Krumholz}
  et~al.}{2006}]{krumholz.matzner.mckee:2006}
{Krumholz} M.~R.,  {Matzner} C.~D.,   {McKee} C.~F.,  2006, \mn@doi [\apj]
  {10.1086/508679}, \href {http://adsabs.harvard.edu/abs/2006ApJ...653..361K}
  {653, 361}

\bibitem[\protect\citeauthoryear{{Krumholz}, {Klein}, {McKee}, {Offner}  \&
  {Cunningham}}{{Krumholz} et~al.}{2009}]{krumholz:2009.massive.sf}
{Krumholz} M.~R.,  {Klein} R.~I.,  {McKee} C.~F.,  {Offner} S.~S.~R.,
  {Cunningham} A.~J.,  2009, \mn@doi [Science] {10.1126/science.1165857}, \href
  {http://adsabs.harvard.edu/abs/2009Sci...323..754K} {323, 754}

\bibitem[\protect\citeauthoryear{{Krumholz}, {Dekel}  \& {McKee}}{{Krumholz}
  et~al.}{2012a}]{krumholz:2012.universal.sf.efficiency}
{Krumholz} M.~R.,  {Dekel} A.,   {McKee} C.~F.,  2012a, \mn@doi [\apj]
  {10.1088/0004-637X/745/1/69}, \href
  {http://adsabs.harvard.edu/abs/2012ApJ...745...69K} {745, 69}

\bibitem[\protect\citeauthoryear{{Krumholz}, {Klein}  \& {McKee}}{{Krumholz}
  et~al.}{2012b}]{krumholz:2012.orion.rhd}
{Krumholz} M.~R.,  {Klein} R.~I.,   {McKee} C.~F.,  2012b, \mn@doi [\apj]
  {10.1088/0004-637X/754/1/71}, \href
  {http://adsabs.harvard.edu/abs/2012ApJ...754...71K} {754, 71}

\bibitem[\protect\citeauthoryear{{Krumholz} et~al.,}{{Krumholz}
  et~al.}{2014}]{krumholz:2014.feedback.review}
{Krumholz} M.~R.,  et~al., 2014, \mn@doi [Protostars and Planets VI]
  {10.2458/azu_uapress_9780816531240-ch011}, \href
  {http://adsabs.harvard.edu/abs/2014prpl.conf..243K} {pp 243--266}

\bibitem[\protect\citeauthoryear{{Lada}, {Margulis}  \& {Dearborn}}{{Lada}
  et~al.}{1984}]{lada:1984}
{Lada} C.~J.,  {Margulis} M.,   {Dearborn} D.,  1984, \mn@doi [\apj]
  {10.1086/162485}, \href {http://adsabs.harvard.edu/abs/1984ApJ...285..141L}
  {285, 141}

\bibitem[\protect\citeauthoryear{{Lada}, {Lombardi}  \& {Alves}}{{Lada}
  et~al.}{2010}]{lada:2010.gmcs}
{Lada} C.~J.,  {Lombardi} M.,   {Alves} J.~F.,  2010, \mn@doi [\apj]
  {10.1088/0004-637X/724/1/687}, \href
  {http://adsabs.harvard.edu/abs/2010ApJ...724..687L} {724, 687}

\bibitem[\protect\citeauthoryear{{Larson}}{{Larson}}{1981}]{larson:gmc.scalings}
{Larson} R.~B.,  1981, \mnras, \href
  {http://adsabs.harvard.edu/abs/1981MNRAS.194..809L} {194, 809}

\bibitem[\protect\citeauthoryear{{Lee}, {Murray}  \& {Rahman}}{{Lee}
  et~al.}{2012}]{lee:2012.wmap}
{Lee} E.~J.,  {Murray} N.,   {Rahman} M.,  2012, \mn@doi [\apj]
  {10.1088/0004-637X/752/2/146}, \href
  {http://adsabs.harvard.edu/abs/2012ApJ...752..146L} {752, 146}

\bibitem[\protect\citeauthoryear{{Lee}, {Chang}  \& {Murray}}{{Lee}
  et~al.}{2015}]{lee:2015.gravoturbulence}
{Lee} E.~J.,  {Chang} P.,   {Murray} N.,  2015, \mn@doi [\apj]
  {10.1088/0004-637X/800/1/49}, \href
  {http://adsabs.harvard.edu/abs/2015ApJ...800...49L} {800, 49}

\bibitem[\protect\citeauthoryear{{Lee}, {Miville-Desch{\^e}nes}  \&
  {Murray}}{{Lee} et~al.}{2016}]{lee:2016.gmc.eff}
{Lee} E.~J.,  {Miville-Desch{\^e}nes} M.-A.,   {Murray} N.~W.,  2016, \mn@doi
  [\apj] {10.3847/1538-4357/833/2/229}, \href
  {http://adsabs.harvard.edu/abs/2016ApJ...833..229L} {833, 229}

\bibitem[\protect\citeauthoryear{{Lombardi}}{{Lombardi}}{2009}]{2009A&A...493..735L}
{Lombardi} M.,  2009, \mn@doi [\aap] {10.1051/0004-6361:200810519}, \href
  {https://ui.adsabs.harvard.edu/#abs/2009A&A...493..735L} {493, 735}

\bibitem[\protect\citeauthoryear{{Lombardi}, {Bouy}, {Alves}  \&
  {Lada}}{{Lombardi} et~al.}{2014}]{lombardi:2014.density.pdf}
{Lombardi} M.,  {Bouy} H.,  {Alves} J.,   {Lada} C.~J.,  2014, \mn@doi [\aap]
  {10.1051/0004-6361/201323293}, \href
  {http://adsabs.harvard.edu/abs/2014A%26A...566A..45L} {566, A45}

\bibitem[\protect\citeauthoryear{{Mathieu}}{{Mathieu}}{1983}]{mathieu:1983}
{Mathieu} R.~D.,  1983, \mn@doi [\apjl] {10.1086/184011}, \href
  {http://adsabs.harvard.edu/abs/1983ApJ...267L..97M} {267, L97}

\bibitem[\protect\citeauthoryear{{McKee} \& {Ostriker}}{{McKee} \&
  {Ostriker}}{2007}]{mckee:2007.review}
{McKee} C.~F.,  {Ostriker} E.~C.,  2007, \mn@doi [\araa]
  {10.1146/annurev.astro.45.051806.110602}, \href
  {http://adsabs.harvard.edu/abs/2007ARA%26A..45..565M} {45, 565}

\bibitem[\protect\citeauthoryear{{McKee} \& {Tan}}{{McKee} \&
  {Tan}}{2003}]{mckee.tan:2003}
{McKee} C.~F.,  {Tan} J.~C.,  2003, \mn@doi [\apj] {10.1086/346149}, \href
  {http://adsabs.harvard.edu/abs/2003ApJ...585..850M} {585, 850}

\bibitem[\protect\citeauthoryear{{McKee} \& {Williams}}{{McKee} \&
  {Williams}}{1997}]{mckee:1997.ob}
{McKee} C.~F.,  {Williams} J.~P.,  1997, \mn@doi [\apj] {10.1086/303587}, \href
  {http://adsabs.harvard.edu/abs/1997ApJ...476..144M} {476, 144}

\bibitem[\protect\citeauthoryear{{Miville-Desch{\^e}nes}, {Murray}  \&
  {Lee}}{{Miville-Desch{\^e}nes} et~al.}{2017}]{miville:2017.gmcs}
{Miville-Desch{\^e}nes} M.-A.,  {Murray} N.,   {Lee} E.~J.,  2017, \mn@doi
  [\apj] {10.3847/1538-4357/834/1/57}, \href
  {http://adsabs.harvard.edu/abs/2017ApJ...834...57M} {834, 57}

\bibitem[\protect\citeauthoryear{{Mooney} \& {Solomon}}{{Mooney} \&
  {Solomon}}{1988}]{mooney.solomon:1988}
{Mooney} T.~J.,  {Solomon} P.~M.,  1988, \mn@doi [\apjl] {10.1086/185310},
  \href {http://adsabs.harvard.edu/abs/1988ApJ...334L..51M} {334, L51}

\bibitem[\protect\citeauthoryear{{Murray}}{{Murray}}{2011}]{murray:2010.sfe.mw.gmc}
{Murray} N.,  2011, \mn@doi [\apj] {10.1088/0004-637X/729/2/133}, \href
  {http://adsabs.harvard.edu/abs/2010arXiv1007.3270M} {729, 133}

\bibitem[\protect\citeauthoryear{{Murray} \& {Chang}}{{Murray} \&
  {Chang}}{2015}]{murray.chang:2015}
{Murray} N.,  {Chang} P.,  2015, \mn@doi [\apj] {10.1088/0004-637X/804/1/44},
  \href {http://adsabs.harvard.edu/abs/2015ApJ...804...44M} {804, 44}

\bibitem[\protect\citeauthoryear{{Murray} \& {Rahman}}{{Murray} \&
  {Rahman}}{2010}]{murray:2010.wmap}
{Murray} N.,  {Rahman} M.,  2010, \mn@doi [\apj] {10.1088/0004-637X/709/1/424},
  \href {http://adsabs.harvard.edu/abs/2010ApJ...709..424M} {709, 424}

\bibitem[\protect\citeauthoryear{{Murray}, {Quataert}  \& {Thompson}}{{Murray}
  et~al.}{2010}]{murray:molcloud.disrupt.by.rad.pressure}
{Murray} N.,  {Quataert} E.,   {Thompson} T.~A.,  2010, \mn@doi [\apj]
  {10.1088/0004-637X/709/1/191}, \href
  {http://adsabs.harvard.edu/abs/2009arXiv0906.5358M} {709, 191}

\bibitem[\protect\citeauthoryear{{Murray}, {Chang}, {Murray}  \&
  {Pittman}}{{Murray} et~al.}{2017}]{murray:2017}
{Murray} D.~W.,  {Chang} P.,  {Murray} N.~W.,   {Pittman} J.,  2017, \mn@doi
  [\mnras] {10.1093/mnras/stw2796}, \href
  {http://adsabs.harvard.edu/abs/2017MNRAS.465.1316M} {465, 1316}

\bibitem[\protect\citeauthoryear{{Myers}, {Dame}, {Thaddeus}, {Cohen},
  {Silverberg}, {Dwek}  \& {Hauser}}{{Myers} et~al.}{1986}]{myers:1986.gmcs}
{Myers} P.~C.,  {Dame} T.~M.,  {Thaddeus} P.,  {Cohen} R.~S.,  {Silverberg}
  R.~F.,  {Dwek} E.,   {Hauser} M.~G.,  1986, \mn@doi [\apj] {10.1086/163909},
  \href {http://adsabs.harvard.edu/abs/1986ApJ...301..398M} {301, 398}

\bibitem[\protect\citeauthoryear{{Myers}, {Klein}, {Krumholz}  \&
  {McKee}}{{Myers} et~al.}{2014a}]{2014MNRAS.439.3420M}
{Myers} A.~T.,  {Klein} R.~I.,  {Krumholz} M.~R.,   {McKee} C.~F.,  2014a,
  \mn@doi [\mnras] {10.1093/mnras/stu190}, \href
  {http://adsabs.harvard.edu/abs/2014MNRAS.439.3420M} {439, 3420}

\bibitem[\protect\citeauthoryear{{Myers}, {Klein}, {Krumholz}  \&
  {McKee}}{{Myers} et~al.}{2014b}]{myers:2014.feedback}
{Myers} A.~T.,  {Klein} R.~I.,  {Krumholz} M.~R.,   {McKee} C.~F.,  2014b,
  \mn@doi [\mnras] {10.1093/mnras/stu190}, \href
  {http://adsabs.harvard.edu/abs/2014MNRAS.439.3420M} {439, 3420}

\bibitem[\protect\citeauthoryear{{Ochsendorf}, {Meixner}, {Roman-Duval},
  {Rahman}  \& {Evans}}{{Ochsendorf} et~al.}{2017}]{ochsendorf:2017.gmcs}
{Ochsendorf} B.~B.,  {Meixner} M.,  {Roman-Duval} J.,  {Rahman} M.,   {Evans}
  II N.~J.,  2017, \mn@doi [\apj] {10.3847/1538-4357/aa704a}, \href
  {http://adsabs.harvard.edu/abs/2017ApJ...841..109O} {841, 109}

\bibitem[\protect\citeauthoryear{{Offner}, {Clark}, {Hennebelle}, {Bastian},
  {Bate}, {Hopkins}, {Moraux}  \& {Whitworth}}{{Offner}
  et~al.}{2014}]{offner:2014.imf.review}
{Offner} S.~S.~R.,  {Clark} P.~C.,  {Hennebelle} P.,  {Bastian} N.,  {Bate}
  M.~R.,  {Hopkins} P.~F.,  {Moraux} E.,   {Whitworth} A.~P.,  2014, \mn@doi
  [Protostars and Planets VI] {10.2458/azu_uapress_9780816531240-ch003}, \href
  {http://adsabs.harvard.edu/abs/2014prpl.conf...53O} {pp 53--75}

\bibitem[\protect\citeauthoryear{{Onus}, {Krumholz}  \& {Federrath}}{{Onus}
  et~al.}{2018}]{onus:2018.hcn}
{Onus} A.,  {Krumholz} M.~R.,   {Federrath} C.,  2018, \mn@doi [\mnras]
  {10.1093/mnras/sty1662}, \href
  {http://adsabs.harvard.edu/abs/2018MNRAS.479.1702O} {479, 1702}

\bibitem[\protect\citeauthoryear{{Orr} et~al.,}{{Orr}
  et~al.}{2018}]{orr:2018.kennicutt.schmidt}
{Orr} M.~E.,  et~al., 2018, \mn@doi [\mnras] {10.1093/mnras/sty1241}, \href
  {http://adsabs.harvard.edu/abs/2018MNRAS.478.3653O} {478, 3653}

\bibitem[\protect\citeauthoryear{{Ostriker} \& {Shetty}}{{Ostriker} \&
  {Shetty}}{2011}]{ostriker.shetty:2011}
{Ostriker} E.~C.,  {Shetty} R.,  2011, \mn@doi [\apj]
  {10.1088/0004-637X/731/1/41}, \href
  {http://adsabs.harvard.edu/abs/2011ApJ...731...41O} {731, 41}

\bibitem[\protect\citeauthoryear{{Padoan} \& {Nordlund}}{{Padoan} \&
  {Nordlund}}{2011}]{pn2011}
{Padoan} P.,  {Nordlund} {\AA}.,  2011, \mn@doi [\apj]
  {10.1088/0004-637X/730/1/40}, \href
  {http://adsabs.harvard.edu/abs/2011ApJ...730...40P} {730, 40}

\bibitem[\protect\citeauthoryear{{Padoan}, {Haugb{\o}lle}  \&
  {Nordlund}}{{Padoan} et~al.}{2012}]{padoan:2012.sfe}
{Padoan} P.,  {Haugb{\o}lle} T.,   {Nordlund} {\AA}.,  2012, \mn@doi [\apjl]
  {10.1088/2041-8205/759/2/L27}, \href
  {http://adsabs.harvard.edu/abs/2012ApJ...759L..27P} {759, L27}

\bibitem[\protect\citeauthoryear{{Palla} \& {Stahler}}{{Palla} \&
  {Stahler}}{2000}]{palla2000}
{Palla} F.,  {Stahler} S.~W.,  2000, \mn@doi [\apj] {10.1086/309312}, \href
  {http://adsabs.harvard.edu/abs/2000ApJ...540..255P} {540, 255}

\bibitem[\protect\citeauthoryear{P\'erez \& Granger}{P\'erez \&
  Granger}{2007}]{ipython}
P\'erez F.,  Granger B.~E.,  2007, \mn@doi [Computing in Science and
  Engineering] {10.1109/MCSE.2007.53}, 9, 21

\bibitem[\protect\citeauthoryear{{Raskutti}, {Ostriker}  \&
  {Skinner}}{{Raskutti} et~al.}{2016}]{raskutti:2016.gmcs}
{Raskutti} S.,  {Ostriker} E.~C.,   {Skinner} M.~A.,  2016, \mn@doi [\apj]
  {10.3847/0004-637X/829/2/130}, \href
  {http://adsabs.harvard.edu/abs/2016ApJ...829..130R} {829, 130}

\bibitem[\protect\citeauthoryear{{Rosen}, {Krumholz}, {McKee}  \&
  {Klein}}{{Rosen} et~al.}{2016}]{rosen:2016.massive.sf}
{Rosen} A.~L.,  {Krumholz} M.~R.,  {McKee} C.~F.,   {Klein} R.~I.,  2016,
  \mn@doi [\mnras] {10.1093/mnras/stw2153}, \href
  {http://adsabs.harvard.edu/abs/2016MNRAS.463.2553R} {463, 2553}

\bibitem[\protect\citeauthoryear{{Rosolowsky}, {Pineda}, {Kauffmann}  \&
  {Goodman}}{{Rosolowsky} et~al.}{2008}]{rosolowsky:2008.gmcs}
{Rosolowsky} E.~W.,  {Pineda} J.~E.,  {Kauffmann} J.,   {Goodman} A.~A.,  2008,
  \mn@doi [\apj] {10.1086/587685}, \href
  {http://adsabs.harvard.edu/abs/2008ApJ...679.1338R} {679, 1338}

\bibitem[\protect\citeauthoryear{{Schneider} et~al.,}{{Schneider}
  et~al.}{2015a}]{schneider:2015a.power.law.tail}
{Schneider} N.,  et~al., 2015a, \mn@doi [\mnras] {10.1093/mnrasl/slv101}, \href
  {http://adsabs.harvard.edu/abs/2015MNRAS.453L..41S} {453, L41}

\bibitem[\protect\citeauthoryear{{Schneider} et~al.,}{{Schneider}
  et~al.}{2015b}]{schneider:2015b.power.law.tail}
{Schneider} N.,  et~al., 2015b, \mn@doi [\aap] {10.1051/0004-6361/201424375},
  \href {http://adsabs.harvard.edu/abs/2015A%26A...578A..29S} {578, A29}

\bibitem[\protect\citeauthoryear{{Scoville} \& {Good}}{{Scoville} \&
  {Good}}{1989}]{scoville:1989.gmcs}
{Scoville} N.~Z.,  {Good} J.~C.,  1989, \mn@doi [\apj] {10.1086/167283}, \href
  {http://adsabs.harvard.edu/abs/1989ApJ...339..149S} {339, 149}

\bibitem[\protect\citeauthoryear{{Shu}, {Adams}  \& {Lizano}}{{Shu}
  et~al.}{1987}]{shu:1987.review}
{Shu} F.~H.,  {Adams} F.~C.,   {Lizano} S.,  1987, \mn@doi [\araa]
  {10.1146/annurev.aa.25.090187.000323}, \href
  {http://adsabs.harvard.edu/abs/1987ARA%26A..25...23S} {25, 23}

\bibitem[\protect\citeauthoryear{{Skinner} \& {Ostriker}}{{Skinner} \&
  {Ostriker}}{2015}]{skinner:2015.ir.molcloud.disrupt}
{Skinner} M.~A.,  {Ostriker} E.~C.,  2015, \mn@doi [\apj]
  {10.1088/0004-637X/809/2/187}, \href
  {http://adsabs.harvard.edu/abs/2015ApJ...809..187S} {809, 187}

\bibitem[\protect\citeauthoryear{{Solomon}, {Rivolo}, {Barrett}  \&
  {Yahil}}{{Solomon} et~al.}{1987}]{solomon:gmc.scalings}
{Solomon} P.~M.,  {Rivolo} A.~R.,  {Barrett} J.,   {Yahil} A.,  1987, \mn@doi
  [\apj] {10.1086/165493}, \href
  {http://adsabs.harvard.edu/abs/1987ApJ...319..730S} {319, 730}

\bibitem[\protect\citeauthoryear{{Sormani}, {Tre{\ss}}, {Klessen}  \&
  {Glover}}{{Sormani} et~al.}{2017}]{sormani:2016.imf.sampling}
{Sormani} M.~C.,  {Tre{\ss}} R.~G.,  {Klessen} R.~S.,   {Glover} S.~C.~O.,
  2017, \mn@doi [\mnras] {10.1093/mnras/stw3205}, \href
  {http://adsabs.harvard.edu/abs/2017MNRAS.466..407S} {466, 407}

\bibitem[\protect\citeauthoryear{{Su} et~al.,}{{Su}
  et~al.}{2018}]{su:2017.discreteness}
{Su} K.-Y.,  et~al., 2018, \mn@doi [\mnras] {10.1093/mnras/sty1928}, \href
  {http://adsabs.harvard.edu/abs/2018MNRAS.480.1666S} {480, 1666}

\bibitem[\protect\citeauthoryear{{Tan}, {Beltr{\'a}n}, {Caselli}, {Fontani},
  {Fuente}, {Krumholz}, {McKee}  \& {Stolte}}{{Tan}
  et~al.}{2014}]{tan:2014.massive.sf}
{Tan} J.~C.,  {Beltr{\'a}n} M.~T.,  {Caselli} P.,  {Fontani} F.,  {Fuente} A.,
  {Krumholz} M.~R.,  {McKee} C.~F.,   {Stolte} A.,  2014, \mn@doi [Protostars
  and Planets VI] {10.2458/azu_uapress_9780816531240-ch007}, \href
  {http://adsabs.harvard.edu/abs/2014prpl.conf..149T} {pp 149--172}

\bibitem[\protect\citeauthoryear{{Thompson} \& {Krumholz}}{{Thompson} \&
  {Krumholz}}{2016}]{thompson:2016.eddington.outflows}
{Thompson} T.~A.,  {Krumholz} M.~R.,  2016, \mn@doi [\mnras]
  {10.1093/mnras/stv2331}, \href
  {http://adsabs.harvard.edu/abs/2016MNRAS.455..334T} {455, 334}

\bibitem[\protect\citeauthoryear{{Thompson}, {Quataert}  \&
  {Murray}}{{Thompson} et~al.}{2005}]{thompson:rad.pressure}
{Thompson} T.~A.,  {Quataert} E.,   {Murray} N.,  2005, \mn@doi [\apj]
  {10.1086/431923}, \href {http://adsabs.harvard.edu/abs/2005ApJ...630..167T}
  {630, 167}

\bibitem[\protect\citeauthoryear{{Tsz-Ho Tsang} \& {Milosavljevic}}{{Tsz-Ho
  Tsang} \& {Milosavljevic}}{2017}]{tsang:2017.ssc.rp}
{Tsz-Ho Tsang} B.,  {Milosavljevic} M.,  2017, preprint, \href
  {http://adsabs.harvard.edu/abs/2017arXiv170907539T} {} (\mn@eprint {arXiv}
  {1709.07539})

\bibitem[\protect\citeauthoryear{{Tutukov}}{{Tutukov}}{1978}]{tutukov:1978}
{Tutukov} A.~V.,  1978, \aap, \href
  {http://adsabs.harvard.edu/abs/1978A%26A....70...57T} {70, 57}

\bibitem[\protect\citeauthoryear{{Vazquez-Semadeni}}{{Vazquez-Semadeni}}{2015}]{vazquez:2015}
{Vazquez-Semadeni} E.,  2015, IAU General Assembly, \href
  {http://adsabs.harvard.edu/abs/2015IAUGA..2250878V} {22, 2250878}

\bibitem[\protect\citeauthoryear{V\'{a}zquez-Semadeni, Col\'{i}n, G\'{o}mez,
  Ballesteros-Paredes  \& Watson}{V\'{a}zquez-Semadeni
  et~al.}{2010}]{vazquez:2010}
V\'{a}zquez-Semadeni E.,  Col\'{i}n P.,  G\'{o}mez G.~C.,  Ballesteros-Paredes
  J.,   Watson A.~W.,  2010, The Astrophysical Journal, 715, 1302

\bibitem[\protect\citeauthoryear{{V{\'a}zquez-Semadeni},
  {Gonz{\'a}lez-Samaniego}  \& {Col{\'{\i}}n}}{{V{\'a}zquez-Semadeni}
  et~al.}{2017}]{vazquez:2017}
{V{\'a}zquez-Semadeni} E.,  {Gonz{\'a}lez-Samaniego} A.,   {Col{\'{\i}}n} P.,
  2017, \mn@doi [\mnras] {10.1093/mnras/stw3229}, \href
  {http://adsabs.harvard.edu/abs/2017MNRAS.467.1313V} {467, 1313}

\bibitem[\protect\citeauthoryear{{Vutisalchavakul}, {Evans}  \&
  {Heyer}}{{Vutisalchavakul} et~al.}{2016}]{vuti:2016.gmcs}
{Vutisalchavakul} N.,  {Evans} II N.~J.,   {Heyer} M.,  2016, \mn@doi [\apj]
  {10.3847/0004-637X/831/1/73}, \href
  {http://adsabs.harvard.edu/abs/2016ApJ...831...73V} {831, 73}

\bibitem[\protect\citeauthoryear{{Williams} \& {McKee}}{{Williams} \&
  {McKee}}{1997}]{williams.mckee:1997}
{Williams} J.~P.,  {McKee} C.~F.,  1997, \mn@doi [\apj] {10.1086/303588}, \href
  {http://adsabs.harvard.edu/abs/1997ApJ...476..166W} {476, 166}

\bibitem[\protect\citeauthoryear{{Wu}, {Evans}, {Gao}, {Solomon}, {Shirley}  \&
  {Vanden Bout}}{{Wu} et~al.}{2005}]{wu:2005.clumps}
{Wu} J.,  {Evans} II N.~J.,  {Gao} Y.,  {Solomon} P.~M.,  {Shirley} Y.~L.,
  {Vanden Bout} P.~A.,  2005, \mn@doi [\apjl] {10.1086/499623}, \href
  {http://adsabs.harvard.edu/abs/2005ApJ...635L.173W} {635, L173}

\bibitem[\protect\citeauthoryear{{Wu}, {Evans}, {Shirley}  \& {Knez}}{{Wu}
  et~al.}{2010}]{wu:2010.clumps}
{Wu} J.,  {Evans} II N.~J.,  {Shirley} Y.~L.,   {Knez} C.,  2010, \mn@doi
  [\apjs] {10.1088/0067-0049/188/2/313}, \href
  {http://adsabs.harvard.edu/abs/2010ApJS..188..313W} {188, 313}

\bibitem[\protect\citeauthoryear{{Zhang} \& {Davis}}{{Zhang} \&
  {Davis}}{2017}]{zhang:2017.ir.pressure}
{Zhang} D.,  {Davis} S.~W.,  2017, \mn@doi [\apj] {10.3847/1538-4357/aa6935},
  \href {http://adsabs.harvard.edu/abs/2017ApJ...839...54Z} {839, 54}

\bibitem[\protect\citeauthoryear{{Zinnecker} \& {Yorke}}{{Zinnecker} \&
  {Yorke}}{2007}]{zinnecker:2007.massive.sf}
{Zinnecker} H.,  {Yorke} H.~W.,  2007, \mn@doi [\araa]
  {10.1146/annurev.astro.44.051905.092549}, \href
  {http://adsabs.harvard.edu/abs/2007ARA%26A..45..481Z} {45, 481}

\bibitem[\protect\citeauthoryear{{Zuckerman} \& {Evans}}{{Zuckerman} \&
  {Evans}}{1974}]{zuckerman1974}
{Zuckerman} B.,  {Evans} II N.~J.,  1974, \mn@doi [\apjl] {10.1086/181613},
  \href {http://adsabs.harvard.edu/abs/1974ApJ...192L.149Z} {192, L149}

\makeatother
\end{thebibliography}







\bsp	
\label{lastpage}
\end{document}